\documentclass[aps,prd,twocolumn,superscriptaddress,nofootinbib,bibliography]{revtex4-2}

\usepackage[T1]{fontenc}
\usepackage[utf8]{inputenc}
\usepackage{lmodern}
\usepackage{amsmath,amssymb,mathtools,bm}
\usepackage{microtype}
\usepackage{tikz}
\usetikzlibrary{calc,arrows.meta,positioning}
\usepackage{comment}
\usepackage[colorlinks=true,citecolor=blue,linkcolor=blue,urlcolor=blue]{hyperref}
\allowdisplaybreaks

\newcommand{\dd}{\mathrm d}
\newcommand{\D}{\nabla}
\newcommand{\G}{\mathcal G}
\newcommand{\T}{\mathcal T}
\newcommand{\eps}{\varepsilon}
\newcommand{\sech}{\operatorname{sech}}
\newcommand{\J}{\mathcal J}
\newcommand{\E}{\mathcal E}
\newcommand{\B}{\mathcal{B}}
\newcommand{\Paux}{\mathcal P}
\newcommand{\gaux}{\mathfrak g}

\definecolor{lime}{HTML}{A6CE39}
\DeclareRobustCommand{\orcidicon}{%
 \begin{tikzpicture}
 \draw[lime, fill=lime] (0,0) 
 circle [radius=0.16] 
 node[white] {{\fontfamily{qag}\selectfont \tiny ID}};
 \draw[white, fill=white] (-0.0625,0.095) 
 circle [radius=0.007];
 \end{tikzpicture}
 \hspace{-2mm}
}

\foreach \x in {A, ..., Z}{%
 \expandafter\xdef\csname orcid\x\endcsname{\noexpand\href{https://orcid.org/\csname orcidauthor\x\endcsname}{\noexpand\orcidicon}}
}

\newcommand\orcidDavid{{\href{https://orcid.org/0009-0006-8485-3140}{\orcidicon}}}
\newcommand\orcidFrancisco{{\href{https://orcid.org/0000-0002-9388-8373}{\orcidicon}}}
\newcommand\orcidJPedro{{\href{https://orcid.org/0000-0002-9758-3366}{\orcidicon}}}

\begin{document}

\title{A Buchdahl solution generator for Einstein--sigma gravity\\ and multiscalar--tensor gravity}

\author{David S. Pereira\orcidDavid\!\!}
\email{djpereira@ciencias.ulisboa.pt}

\author{Francisco S.N. Lobo\orcidFrancisco\!\!}
\email{fslobo@ciencias.ulisboa.pt}

\author{Jos\'e Pedro Mimoso\orcidJPedro\!\!}
\email{jpmimoso@ciencias.ulisboa.pt}

\affiliation{Departamento de F\'{i}sica, Faculdade de Ci\^{e}ncias da Universidade de Lisboa, Campo Grande, Edif\'{i}cio C8, P-1749-016 Lisbon, Portugal}
\affiliation{Instituto de Astrof\'{i}sica e Ci\^{e}ncias do Espa\c{c}o, Faculdade de Ci\^{e}ncias da Universidade de Lisboa, Campo Grande, P-1749-016 Lisbon, Portugal}

\date{\today}

\begin{abstract}
Buchdahl transformations provide a geometric route from vacuum spacetimes to scalar-field solutions. We formulate this mechanism as a target-covariant solution generator for Einstein gravity coupled to nonlinear sigma models in $D>3$ dimensions. On any connected region in which a Ricci-flat seed admits a non-null hypersurface-orthogonal cyclic coordinate with fixed-sign nonzero norm, the logarithm of that norm defines a Buchdahl potential $\Sigma$. A one-parameter redistribution of the cyclic norm then generates the exact Ricci source $R_{AB}=C_D(\beta)\partial_A\Sigma\partial_B\Sigma$, with $C_D(\beta)=\frac{D-2}{D-3}(1-\beta^2)$, while $\Sigma$ remains harmonic in the deformed metric. Composing $\Sigma$ with an affinely parametrized geodesic of a non-degenerate target metric whose squared speed is $C_D(\beta)$ therefore yields an exact Einstein--sigma-model solution. For positive-definite targets, the corresponding pullback condition also implies that the nontrivial scalar map has one-dimensional image; for indefinite targets, the geodesic ansatz remains a sufficient solution-generating sector but need not exhaust all possible realizations of the same rank-one spacetime source. Through the Einstein/Jordan-frame correspondence, the construction gives vacuum tensor--multiscalar solutions as well. We illustrate the unified generator with flat, spherical, hyperbolic and axion--dilaton targets and with Tangherlini, Weyl, Kasner, Rosen-wave, Gowdy, Kaluza--Klein bubble and C-metric seeds. The result separates cleanly the geometric Buchdahl source from its target-space realization and extends this rank-one cyclic sector across static, cosmological, wave and accelerating geometries.
\end{abstract}

\maketitle

\section{Introduction}

Exact solutions occupy a special place in gravitational physics. They are not only examples of the Einstein equations; they are controlled laboratories in which one can identify invariant singularities, characterize horizons, define asymptotic charges, test no-hair statements, understand causal structure, and separate genuine physical effects from coordinate artifacts. The basic solutions of general relativity have shaped the interpretation of the theory: Schwarzschild and Kerr geometries define the black-hole paradigm, Reissner--Nordstr\"om and Kerr--Newman solutions show how additional long-range fields alter horizons, Kasner and Bianchi metrics reveal anisotropic singularity structure, plane waves and Einstein--Rosen waves display exact radiative degrees of freedom, Gowdy metrics provide inhomogeneous compact cosmologies, and the C-metric describes uniformly accelerated sources. For these reasons, catalogues and solution-generating methods are not merely technical exercises; they expose hidden structures of the field equations and provide benchmarks for perturbative, numerical and effective descriptions of gravity~\cite{Wald:1984rg,Stephani:2003tm,GriffithsPodolsky2009, MacCallum:2006mf,Bekenstein:1974sf}.

Scalar fields are among the most important matter sectors in which such exact solutions can be tested. They appear in scalar--tensor gravity, from the Brans--Dicke theory to modern tensor--scalar and tensor--multiscalar models, as well as in Kaluza--Klein reductions, low-energy string actions, supergravity, moduli dynamics and cosmological effective theories ~\cite{Brans:1961sx,Bergmann:1968ve,Wagoner:1970vr,Damour:1992we, Damour:1993hw,Faraoni:2004pi,Gibbons:1987ps, Garfinkle:1990qj,Maharana:1992my,Kallosh:1994ba}. Even a single massless scalar can change the global structure of a vacuum spacetime: in the static spherically symmetric sector, the Fisher--Janis--Newman--Winicour--Wyman geometry shows how scalar charge replaces the Schwarzschild horizon by a curvature singularity \cite{Fisher:1948yn,Janis:1968zz,Wyman:1981bd,Wiltshire:1990ah, Wiltshire:1992jc}. This behaviour is closely related to classical no-scalar-hair results for asymptotically flat black holes \cite{Chase:1970omy,Bekenstein:1995un,Herdeiro:2015waa,Reiris:2015zaa,Sotiriou:2011dz,Hawking:1972qk,Herdeiro:2023mpt}. Thus scalar exact solutions are useful both as explicit matter configurations and as probes of the assumptions behind black-hole uniqueness and no-hair theorems.

In many theories, however, the scalar sector is not naturally a single canonical field. Several scalars often appear together, and their kinetic terms define a nonlinear sigma model. The scalar fields are then coordinates on a target manifold $\mathcal T$, with metric $\mathcal G_{IJ}$, rather than independent unrelated fields. This target-space viewpoint is standard in harmonic-map theory, symmetric-space reductions, Kaluza--Klein and supergravity models, axion--dilaton systems, and tensor--multiscalar gravity \cite{,Ehlers:1959aug,Geroch:1970nt,Eells:1964,Helgason1978,Misner:1978am, Maison:1980rf,Maison:1979kx,Maison:1978rt,Belinskii:1973sud, Breitenlohner:1987dg,Ortin:2004ms,Damour:1992we}. It also has an important physical consequence: the same spacetime metric can carry different scalar interpretations depending on the target geometry and on the trajectory followed by the scalar map.

The Einstein-frame theory considered in the main part of this paper is
\begin{equation}\label{eq:action_D}
S_\E =\frac{1}{2\kappa_D}\int \dd^Dx\sqrt{|g|}\left[R[g]-\G_{IJ}(\varphi)g^{AB}\partial_A\varphi^I\partial_B\varphi^J\right] .
\end{equation}
With this normalization there is no additional factor of $1/2$ multiplying the scalar kinetic term inside the Einstein bracket. The field equations are therefore
\begin{equation}\label{eq:Einstein_sigma_D}
R_{AB}=\G_{IJ}(\varphi)\partial_A\varphi^I\partial_B\varphi^J,
\end{equation}
and
\begin{equation}\label{eq:scalar_sigma_D}
\Box_g\varphi^I+\Gamma^I{}_{JK}(\varphi)g^{AB}\partial_A\varphi^J\partial_B\varphi^K=0 ,
\end{equation}
where $\Gamma^I{}_{JK}$ is the Levi-Civita connection of the target metric. The problem is to construct exact solutions of Eqs.~\eqref{eq:Einstein_sigma_D}--\eqref{eq:scalar_sigma_D} without choosing target coordinates that prematurely reduce the system to a set of decoupled canonical scalars.

Buchdahl's reciprocal method is one of the classical routes from vacuum geometry to scalar-field geometry. In its original form, the method exploits a non-null hypersurface-orthogonal Killing direction and redistributes the logarithm of its norm between that direction and the transverse metric block~\cite{Buchdahl:1956zz,Buchdahl:1959nk,Pereira:2024olv,Mimoso:2011eh}. In the familiar static spherical case this mechanism is related to the scalar geometries of Fisher, Janis--Newman--Winicour and Wyman~\cite{Fisher:1948yn,Janis:1968zz,Wyman:1981bd}. More recent work has emphasized that Buchdahl-type transformations extend well beyond the static spherical setting, including higher dimensions, Einstein--Maxwell--scalar systems, spacelike cyclic directions, rotating or Kerr--Schild constructions, and multipolar scalar geometries~\cite{Matos:1998bd,Matos:1995my,Maeda:2019tqs,Barrientos:2024uuq,Barrientos:2025abs,Lim:2026cky,Campoamor-Stursberg:2024ako}. Thus neither the use of a spacelike cyclic direction nor the existence of non-spherical Buchdahl descendants is, by itself, the novelty sought here.

Exact solutions of multidimensional gravity coupled to nonlinear sigma-model sources have also been constructed using product-manifold and Einstein-factor-space ans\"atze. In particular, Ref.~\cite{Golubtsova:2012hg} obtained classes of cosmological and spherically symmetric solutions in gravity with nonlinear sigma-model sources, with part of the scalar dynamics governed by geodesic equations on the target manifold, and discussed associated no-hair restrictions. This work is close in spirit to the target-space part of the present construction, but the sigma-model source there is introduced through a product-space or warped-factor ansatz. Here, instead, the source is generated geometrically from a Ricci-flat cyclic seed by the Buchdahl deformation itself.

A separate and directly relevant observation is that a known one-scalar Einstein--Klein--Gordon solution may be lifted to a nonlinear sigma model by allowing the scalar fields to move along a geodesic of the target metric~\cite{Brax:2023now}. That target-geometric lifting principle is powerful and applies without requiring target-space isometries. It does not, however, determine the geometric origin of the underlying one-scalar source. The present construction supplies precisely this complementary ingredient: the cyclic vacuum geometry fixes a universal rank-one Ricci source before any target-space realization is chosen.

Accordingly, the novelty of the present work is the unified structural statement obtained by combining these ingredients at the level of the cyclic Einstein--Hilbert density and, more decisively, at the Ricci-tensor level. For any $D>3$ Ricci-flat seed admitting a non-null hypersurface-orthogonal cyclic coordinate on a connected fixed-sign domain, the cyclic norm defines a single Buchdahl potential, and a one-parameter deformation of the cyclic block produces an exact universal source proportional to $\partial_A\Sigma\partial_B\Sigma$. A target-space geodesic then realizes that source in an arbitrary non-degenerate sigma-model target. The spacetime and target-space problems are thereby separated: the seed supplies $\Sigma$, the Buchdahl parameter fixes the invariant scalar strength, and the target geodesic determines the physical scalar fields.

This formulation has three advantages. First, it is dimension independent for $D>3$ and applies locally to timelike or spacelike hypersurface-orthogonal cyclic directions. Second, it is target covariant: flat, compact, hyperbolic, axion--dilaton and more general curved targets are treated in the same language. Third, it gives a precise classification for positive-definite targets: the nontrivial pullback has rank one and hence the scalar map has one-dimensional image. For indefinite targets, null cancellations can make the pullback rank smaller than the differential of the scalar map, so additional degenerate realizations may occur. There is also a distinct spacetime effect when the Buchdahl potential has a null gradient: in that case the scalar equations do not require the one-dimensional target curve to be geodesic, and more general constant-speed curves are admissible. The geodesic construction developed here remains a universal sufficient realization in all of these cases.

The same viewpoint extends to the vacuum gravitational sector of tensor--multiscalar gravity in the Jordan frame. The relevant metric for the target-space geodesic problem is the Einstein-frame target metric obtained after the conformal transformation~\cite{Damour:1992we,Faraoni:1998qx,Faraoni:1999hp,Faraoni:2004pi}. Consequently, two target geodesics with the same speed can support the same Einstein-frame metric while leading to inequivalent Jordan-frame geometries because the physical metric also depends on the position of the geodesic through the nonminimal coupling function. This makes the target-space trajectory physically relevant rather than merely a parametrization.

The examples below illustrate the scope of the generator. In the spherical sector it reproduces the higher-dimensional Tangherlini and four-dimensional Fisher--Janis--Newman--Winicour--Wyman type families \cite{Tangherlini:1963bw,Fisher:1948yn,Janis:1968zz,Wyman:1981bd}. In the static axisymmetric and generalized Weyl sectors it acts on four-dimensional Weyl, Zipoy--Voorhees and Levi-Civita-type seeds, as well as higher-dimensional static geometries with several orthogonal cyclic directions ~\cite{Weyl:1917rtf,Emparan:2001wk,Zipoy:1966btu,Voorhees:1970ywo}. In time-dependent and wave sectors it generates scalar families from Kasner, Rosen and polarized Gowdy seeds \cite{Kasner:1921zz,rosen1937plane,Gowdy:1971jh,Gowdy:1973mu,Moncrief1981,Ringstrom:2009,Ringstrom:2010zz}. It also applies to appropriate Ricci-flat accelerating C-metric patches~\cite{KinnersleyWalker1970,HongTeo2003,GriffithsKrtousPodolsky2006,GriffithsPodolsky2009,AnjomshoaMirzaAzizallahi2025}. In the canonical one-scalar limit, the accelerating branch is closely related to the accelerating FJNW construction of Ref.~\cite{AnjomshoaMirzaAzizallahi2025}; the additional content here is its derivation from the unified cyclic theorem and its extension to arbitrary target-space geodesics and Jordan-frame descendants. These examples show that the Buchdahl potential need not be radial or even purely time dependent; depending on the seed and on the chosen cyclic direction, its gradient can be spacelike, timelike, null or inhomogeneous.

The paper is organized as follows. Section~\ref{sec:method} derives the cyclic curvature-density identity, the Buchdahl curve and the sigma-model completion. Section~\ref{sec:jordan} gives the Jordan-frame tensor--multiscalar extension. Section~\ref{sec:families} presents explicit target geometries and seed families, displaying the generated metrics and scalar profiles. Section \ref{sec:interpretation} discusses the rank-one character, the novelty of the construction and its relation to no-hair expectations. The appendices collect the density derivation and the direct field-equation check. We use the signature $(-,+\ldots+)$, the spacetime dimension is denoted by $D$, uppercase Latin indices are spacetime indices, and lowercase Latin indices refer to the $(D-1)$-dimensional base obtained after isolating one cyclic coordinate.
\section{Multi-field Buchdahl construction}\label{sec:method}

In this section we derive the solution-generating method for the field equations of Eq.~\eqref{eq:action_D}. The full derivation with all the calculations is given in Appendix~\ref{app:cyclic_density_derivation}.

Let $y$ be a non-null hypersurface-orthogonal cyclic coordinate. We consider metrics of the form
\begin{equation}\label{eq:general_cyclic_metric}
\dd s_D^2=\eps e^{2\gamma(x)}\dd y^2+e^{2\chi(x)}h_{ij}(x)\dd x^i\dd x^j ,
\end{equation}
where $\eps=\pm1$, $i,j=1,\ldots,D-1$ not including the $y$ coordinate, and all functions are independent of $y$. The word cyclic means
\begin{equation}\label{eq:cyclic_conditions_D}
\partial_y g_{AB}=0,
\quad
g_{yi}=0,
\quad
g_{yy}\neq0 .
\end{equation}
Thus stationary metrics with off-diagonal terms, such as $g_{t\phi}$, are not included unless they can be put into a hypersurface-orthogonal cyclic block on the domain under consideration. The construction is local and is applied on connected regions where $g_{yy}$ is nonzero and has fixed sign, so that $\eps=\operatorname{sgn}(g_{yy})$ is constant and $\Sigma=\frac12\ln|g_{yy}|$ is smooth. If a seed contains a horizon, axis or other hypersurface at which the cyclic norm vanishes or changes sign, adjacent regions must be treated separately.

The determinant and Ricci scalar of Eq.~\eqref{eq:general_cyclic_metric} combine in a simple way. After discarding total divergences on the base, the Einstein--Hilbert density can be written as
\begin{align}\label{eq:D_density_general}
&\sqrt{|g_D|}R[g_D]
\doteq
\sqrt{|h|}e^F
\Big\{
R[h] \nonumber \\ 
&\qquad + (D-2) \big[ 2\D\gamma\cdot\D\chi +(D-3)(\D\chi)^2 \big] \Big\},
\end{align}
where
\begin{equation}\label{eq:F_def}
F=\gamma+(D-3)\chi .
\end{equation}

All contractions in Eq.~\eqref{eq:D_density_general} are taken with $h_{ij}$, and $\doteq$ denotes equality up to a total divergence. The calculation is local and off shell: no scalar field, no field equation and no vacuum seed has been used.

We now choose the Buchdahl curve in the two functions $\gamma$ and $\chi$:
\begin{equation}\label{eq:buch_split_D}
\gamma=\beta\Sigma,
\quad
\chi=\frac{1-\beta}{D-3}\Sigma,
\quad D>3,
\end{equation}
where $\beta$ is constant. Along this curve $F=\Sigma$, independently of $\beta$. Substitution into Eq.~\eqref{eq:D_density_general} gives
\begin{equation}\label{eq:D_density_beta}
\sqrt{|g_\beta|}R[g_\beta] \doteq \sqrt{|h|}e^\Sigma \left[ R[h]+ \frac{D-2}{D-3}(1-\beta^2)(\D\Sigma)^2 \right],
\end{equation}
where $(\D\Sigma)^2=h^{ij}\D_i\Sigma\D_j\Sigma$. For $\beta=1$, one has
\begin{equation}\label{eq:beta_one_member}
\dd s_1^2= \eps e^{2\Sigma}\dd y^2 + h_{ij}\dd x^i\dd x^j,
\end{equation}
and
\begin{equation}\label{eq:D_seed_density}
\sqrt{|g_1|}R[g_1] \doteq\sqrt{|h|}e^\Sigma R[h] ,
\end{equation}
allowing to write
\begin{equation}\label{eq:D_excess_density_rearranged}
\sqrt{|g_\beta|}R[g_\beta]
-
\sqrt{|h|}e^\Sigma C_D(\beta)(\D\Sigma)^2
\doteq
\sqrt{|g_1|}R[g_1],
\end{equation}
with
\begin{equation}\label{eq:CD_def}
C_D(\beta)=\frac{D-2}{D-3}(1-\beta^2).
\end{equation}

Since $\Sigma$ is independent of the cyclic coordinate $y$, the inverse metric of $g_\beta$ gives
\begin{equation}\label{eq:Sigma_kinetic_full_metric}
g_\beta^{AB}\partial_A\Sigma\partial_B\Sigma
=
e^{2(\beta-1)\Sigma/(D-3)}h^{ij}\D_i\Sigma\D_j\Sigma .
\end{equation}
and using
\begin{equation}
\sqrt{|g_\beta|}=e^{[D-1-2\beta]\Sigma/(D-3)}\sqrt{|h|},
\end{equation}
Eq.~\eqref{eq:D_excess_density_rearranged} can be recast as
\begin{align}\label{eq:key1}
&\int \dd^Dx\,\sqrt{|g_1|}R[g_1]
\doteq
\int \dd^Dx\,\sqrt{|g_\beta|}
\left[
R[g_\beta] \right. \nonumber\\
&\quad \left.-C_D(\beta)g_\beta^{AB}\partial_A\Sigma\partial_B\Sigma \right]
=\int \dd^Dx\,\sqrt{|g_\beta|}
\left[R[g_\beta] \right.\nonumber \\
&\quad\quad \left.-C_D(\beta)e^{2(\beta-1)\Sigma/(D-3)}h^{ij}\D_i\Sigma\D_j\Sigma
\right].
\end{align}

We will now compare the right-hand side of Eq.~\eqref{eq:key1} with the Einstein--sigma-model action~\eqref{eq:action_D}. Before identifying the fields with the physical sigma-model variables, we will introduce $N$ generic scalar variables $\psi^I$, with $I=1,\ldots,N$ expressing the Buchdahl function $\Sigma$ as
\begin{equation}\label{eq:Sigma_as_generic_field_function}
\Sigma(x)=\Sigma\!\left(\psi^1(x),\ldots,\psi^N(x)\right).
\end{equation}
These fields coordinatize an auxiliary field
space $\Paux$. This statement only says that the single Buchdahl degree of freedom may be written as one combination of $N$ variables. It does not specify the physical
target metric.

The chain rule gives
\begin{equation}\label{eq:Sigma_chain_rule_generic_fields}
\partial_A\Sigma
=
\Sigma_{,I}\partial_A\psi^I,
\quad
\Sigma_{,I}\equiv\frac{\partial\Sigma}{\partial\psi^I},
\end{equation}
allowing the Buchdahl kinetic density to be written as
\begin{equation}\label{eq:Buchdahl_rank_one_kinetic_generic}
C_D(\beta)\partial_A\Sigma\partial^A\Sigma = \B_{IJ}(\psi)\partial_A\psi^I\partial^A\psi^J,
\end{equation}
where
\begin{equation}\label{eq:B_rank_one_generic}
\B_{IJ}(\psi)=C_D(\beta)\Sigma_{,I}\Sigma_{,J}.
\end{equation}

The tensor $\B_{IJ}$ is a symmetric $(0,2)$-tensor on $\Paux$. In any coordinate chart it is the outer product of the single covector $\dd\Sigma$ with itself. Therefore, wherever $C_D(\beta)\neq0$ and $\dd\Sigma\neq0$, it has matrix rank one; at the endpoints $C_D(\beta)=0$, or at critical points of $\Sigma$, its rank drops to zero. This is why a single Buchdahl potential cannot produce a generic full-rank $N$-field kinetic metric.

Let $\gaux_{IJ}(\psi)$ be a non-degenerate metric on $\Paux$. We require
the auxiliary sigma-model source to reproduce the Buchdahl source on the field
configuration being constructed:
\begin{equation}\label{eq:matching_general_auxiliary}
\gaux_{IJ}(\psi)\partial_A\psi^I\partial_B\psi^J
=
C_D(\beta)\partial_A\Sigma\partial_B\Sigma .
\end{equation}
This condition is a pullback condition along the scalar map. It does not imply
that $\gaux_{IJ}$ is rank one on the full auxiliary field space.

For positive-definite $\gaux_{IJ}$, Eq.~\eqref{eq:matching_general_auxiliary} implies that the differential of the scalar map has rank at most one wherever the right-hand side is nonzero. Hence the nontrivial scalar map has locally one-dimensional image. Let this image be a curve $\mathcal C_\psi\subset\Paux$, parametrized by $s$:
\begin{equation}\label{eq:curve_scalar_map_auxiliary}
\psi^I(x)=\Psi^I(s(x)).
\end{equation}
Substitution into Eq.~\eqref{eq:matching_general_auxiliary} gives
\begin{equation}\label{eq:curve_matching_tensor_auxiliary}
\gaux_{IJ}(\Psi)
\frac{\dd\Psi^I}{\dd s}
\frac{\dd\Psi^J}{\dd s}
\partial_A s\,\partial_B s
=
C_D(\beta)\partial_A\Sigma\,\partial_B\Sigma .
\end{equation}

Both sides are rank-one spacetime tensors. On each connected region where the
scalar gradient is nonzero, $\partial_A s$ and $\partial_A\Sigma$ are
therefore proportional, so $s=s(\Sigma)$. If the relation is monotonic, one
uses $\Sigma$ itself as the parameter along the curve:
\begin{equation}\label{eq:scalar_curve_sigma_auxiliary}
\psi^I(x)=\Psi^I(\Sigma(x)).
\end{equation}
Then Eq.~\eqref{eq:curve_matching_tensor_auxiliary} reduces to the fixed-speed
condition
\begin{equation}\label{eq:speed_condition_auxiliary}
\gaux_{IJ}(\Psi)
\frac{\dd\Psi^I}{\dd\Sigma}
\frac{\dd\Psi^J}{\dd\Sigma}
=
C_D(\beta).
\end{equation}
For an indefinite target metric, the converse rank statement requires care: a higher-rank scalar-map differential can have a degenerate rank-one pullback because of null directions and cancellations. The geodesic ansatz \eqref{eq:scalar_curve_sigma_auxiliary} therefore defines a universal sufficient realization of the Buchdahl source for any non-degenerate target metric, whereas the statement that the source necessarily forces a one-dimensional target-space image is restricted here to the positive-definite case.

For the metric $\gaux_{IJ}$, the scalar equations have the sigma-model form
\begin{equation}\label{eq:auxiliary_scalar_equation}
\Box_{g_\beta}\psi^I
+
\widehat\Gamma^I{}_{JK}(\psi)g_\beta^{AB}\partial_A\psi^J\partial_B\psi^K=0,
\end{equation}
where $\widehat\Gamma^I{}_{JK}$ is the Levi-Civita connection of
$\gaux_{IJ}$. Using $\psi^I=\Psi^I(\Sigma)$, the chain rule gives
\begin{align}\label{eq:scalar_eq_curve_decomp_auxiliary}
&
\Box_{g_\beta}\psi^I
+
\widehat\Gamma^I{}_{JK}(\Psi)g_\beta^{AB}\partial_A\psi^J\partial_B\psi^K
=\frac{\dd\Psi^I}{\dd\Sigma}\Box_{g_\beta}\Sigma\nonumber\\
&+
\left[
\frac{\dd^2\Psi^I}{\dd\Sigma^2}
+
\widehat\Gamma^I{}_{JK}(\Psi)
\frac{\dd\Psi^J}{\dd\Sigma}
\frac{\dd\Psi^K}{\dd\Sigma}
\right]
g_\beta^{AB}\partial_A\Sigma\partial_B\Sigma .
\end{align}
The effective Buchdahl scalar obeys
\begin{equation}\label{eq:Sigma_wave_equation}
\Box_{g_\beta}\Sigma=0,
\end{equation}
as part of the one-scalar Buchdahl system. Therefore the auxiliary scalar
equations are solved if the curve obeys
\begin{equation}\label{eq:auxiliary_geodesic_equation}
\frac{\dd^2\Psi^I}{\dd\Sigma^2}
+
\widehat\Gamma^I{}_{JK}(\Psi)
\frac{\dd\Psi^J}{\dd\Sigma}
\frac{\dd\Psi^K}{\dd\Sigma}=0,
\end{equation}
and therefore $\mathcal C_\psi$ is an affinely parametrized geodesic of $\gaux_{IJ}$, with affine speed fixed by Eq.~\eqref{eq:speed_condition_auxiliary}. 

At this point the auxiliary field space has served only as bookkeeping for the rank-one pullback. We now identify it with the physical target space by taking
\begin{equation}\label{eq:last_identification}
\gaux_{IJ}(\psi)=\G_{IJ}(\varphi),
\quad
\psi^I=\varphi^I .
\end{equation}
More generally, one may first perform an invertible field redefinition, in which case $\gaux_{IJ}$ is the corresponding pullback of the physical target metric. With the direct identification \eqref{eq:last_identification}, the curve $\mathcal C_\psi$ becomes a curve $\mathcal C\subset\T$ and the physical scalar map is
\begin{equation}\label{eq:physical_curve_solution}
\varphi^I(x)=\Phi^I(\Sigma(x)),
\quad
\Phi^I\equiv\Psi^I.
\end{equation}

Thus, the curve $\mathcal C$ is an affinely parametrized geodesic of $\G_{IJ}$,
with speed
\begin{equation}\label{eq:speed_condition_D}
\G_{IJ}(\Phi)
\frac{\dd\Phi^I}{\dd\Sigma}
\frac{\dd\Phi^J}{\dd\Sigma}=C_D(\beta).
\end{equation}
The physical target curve obeys
\begin{equation}\label{eq:target_geodesic_D}
\frac{\dd^2\Phi^I}{\dd\Sigma^2}
+
\Gamma^I{}_{JK}(\Phi)
\frac{\dd\Phi^J}{\dd\Sigma}
\frac{\dd\Phi^K}{\dd\Sigma}=0.
\end{equation}

For the universal geodesic sector considered here, the curve $\Phi^I(\Sigma)$ is chosen to satisfy the geodesic equation \eqref{eq:target_geodesic_D} and the speed condition \eqref{eq:speed_condition_D} for the target metric $\G_{IJ}$. We now prove directly, at the Ricci-tensor level, that the associated spacetime metric supplies exactly the source required by these scalar fields. 

Let the $\beta=1$ member be the Ricci-flat seed
\begin{equation}\label{eq:seed_metric_D}
\dd s_0^2=\eps e^{2\Sigma}\dd y^2+h_{ij}\dd x^i\dd x^j,
\end{equation}
so that on the chosen fixed-sign domain
\begin{equation}\label{eq:Sigma_sol}
\Sigma=\frac12\ln|g^{(0)}_{yy}|.
\end{equation}
The generated metric is
\begin{equation}\label{eq:D_buch_metric}
\dd s_\beta^2 = \eps e^{2\beta\Sigma}\dd y^2 + e^{2(1-\beta)\Sigma/(D-3)}h_{ij}\dd x^i\dd x^j ,
\end{equation}
or, equivalently,
\begin{equation}\label{eq:D_component_map}
g^{(\beta)}_{yy}=\eps |g^{(0)}_{yy}|^\beta,
\qquad
g^{(\beta)}_{ij}=|g^{(0)}_{yy}|^{(1-\beta)/(D-3)}g^{(0)}_{ij}.
\end{equation}

Ricci flatness of the seed implies
\begin{align}\label{eq:seed_relations_main}
\D^2\Sigma+(\D\Sigma)^2&=0,
\\
R_{ij}[h]&=\D_i\D_j\Sigma+\D_i\Sigma\D_j\Sigma .
\end{align}
Writing
\begin{equation}\label{eq:ak_main}
a=\frac{1-\beta}{D-3},
\qquad
k=\beta-a,
\end{equation}
a direct Ricci calculation gives $R_{yy}[g_\beta]=0$ and
\begin{align}\label{eq:buchdahl_ricci_main}
R_{ij}[g_\beta]
&=\left[1-k^2+(D-2)a^2\right]\D_i\Sigma\D_j\Sigma
\nonumber\\
&=\frac{D-2}{D-3}(1-\beta^2)\D_i\Sigma\D_j\Sigma .
\end{align}
Because $\partial_y\Sigma=0$, the full tensor equation is therefore
\begin{equation}\label{eq:buchdahl_ricci_tensor_main}
R_{AB}[g_\beta]
=C_D(\beta)\partial_A\Sigma\partial_B\Sigma .
\end{equation}
The same calculation yields
\begin{equation}\label{eq:buchdahl_wave_main}
\Box_{g_\beta}\Sigma=0 .
\end{equation}
The detailed curvature calculation is given in Appendix~\ref{app:field_equation_check}.

Now use $\varphi^I=\Phi^I(\Sigma)$. The speed condition \eqref{eq:speed_condition_D} immediately gives
\begin{equation}\label{eq:main_metric_field_check}
\G_{IJ}\partial_A\varphi^I\partial_B\varphi^J
=C_D(\beta)\partial_A\Sigma\partial_B\Sigma
=R_{AB}[g_\beta],
\end{equation}
while the scalar equations decompose as
\begin{align}\label{eq:main_scalar_field_check}
\Box_{g_\beta}\varphi^I
+\Gamma^I{}_{JK}g_\beta^{AB}\partial_A\varphi^J\partial_B\varphi^K
=\frac{\dd\Phi^I}{\dd\Sigma}\Box_{g_\beta}\Sigma
\nonumber\\
+\left[
\frac{\dd^2\Phi^I}{\dd\Sigma^2}
+\Gamma^I{}_{JK}\frac{\dd\Phi^J}{\dd\Sigma}\frac{\dd\Phi^K}{\dd\Sigma}
\right](\nabla\Sigma)^2=0 .
\end{align}
Equation~\eqref{eq:main_scalar_field_check} also makes explicit the role of the causal character of the Buchdahl gradient. On any open region where $(\nabla\Sigma)^2\neq0$, the scalar equations require \eqref{eq:target_geodesic_D} within the ansatz $\varphi^I=\Phi^I(\Sigma)$. If instead $\dd\Sigma\neq0$ is null on an open region, so that $(\nabla\Sigma)^2=0$, the scalar equations impose no target-space geodesic condition. The Einstein equations still enforce the fixed-speed condition \eqref{eq:speed_condition_D}, so more general constant-speed target curves are admissible in this null-gradient sector. We will not pursue this enlargement here and focus in the geodesic subclass, which provides a uniform solution-generating prescription for all seeds. With this choice, the full, unrestricted Einstein--sigma field equations are satisfied directly, without recourse to a restricted variational argument.

\emph{Solution-generating theorem.} Let $g^{(0)}$ be a Ricci-flat seed of the form \eqref{eq:seed_metric_D} on a connected domain where the non-null hypersurface-orthogonal cyclic norm is nonzero and has fixed sign. Let $\Phi^I(\Sigma)$ be an affinely parametrized geodesic of a non-degenerate physical target metric $\G_{IJ}$ satisfying Eq.~\eqref{eq:speed_condition_D}. Then the metric \eqref{eq:D_buch_metric}, together with the scalars \eqref{eq:physical_curve_solution}, solves Eqs.~\eqref{eq:Einstein_sigma_D}--\eqref{eq:scalar_sigma_D}.

The coefficient $C_D(\beta)$ vanishes at both endpoints $\beta=\pm1$. For a positive-definite target metric, the fixed-speed condition forces the target curve to be constant there. The endpoint $\beta=1$ is the original Ricci-flat seed, while $\beta=-1$ is its Buchdahl reciprocal Ricci-flat member; hence only $-1<\beta<1$ gives a nonzero source for a positive-definite target. For an indefinite target, $C_D=0$ may also be realized by a nonzero null tangent. More generally, as emphasized above, the geodesic construction remains sufficient for indefinite targets but is not claimed to exhaust every possible scalar map whose pullback reproduces the same rank-one spacetime tensor.

\section{Jordan-frame tensor--multiscalar extension}\label{sec:jordan}

Many scalar theories are naturally written in a Jordan frame, where matter couples minimally to $\tilde g_{AB}$ and the scalars multiply the Ricci scalar. This includes the Brans--Dicke theory and its tensor--scalar and tensor--multiscalar generalizations, where conformal transformations relate the Jordan-frame and Einstein-frame descriptions \cite{Brans:1961sx,Damour:1992we,Damour:1993hw,
Faraoni:2004pi,Faraoni:1998qx,Faraoni:1999hp}. The Buchdahl--sigma construction extends directly to the vacuum scalar--gravity sector of these theories by using this Einstein/Jordan-frame dictionary. Additional nonvanishing Jordan-frame matter fields, if present, must of course satisfy their own equations and are not generated by the construction below. Consider the vacuum Jordan-frame gravitational action~\cite{Damour:1992we}
\begin{align}\label{eq:jordan_action}
S_\J=&\frac{1}{2\kappa_D}\int\dd^Dx\sqrt{|\tilde g|} \left[ \mathcal A(\varphi)\tilde R  \right. \nonumber\\ 
&\left.-\mathcal H_{IJ}(\varphi) \tilde g^{AB}\partial_A\varphi^I\partial_B\varphi^J \right] ,
\end{align}
where $\mathcal A(\varphi)>0$. The Einstein-frame metric is related to the Jordan-frame metric by
\begin{equation}\label{eq:jordan_to_einstein}
g_{AB}=\mathcal A(\varphi)^{2/(D-2)}\tilde g_{AB},
\end{equation}
or equivalently,
\begin{equation}\label{eq:einstein_to_jordan}
\tilde g_{AB}=\mathcal A(\varphi)^{-2/(D-2)}g_{AB}.
\end{equation}
Up to a boundary term, the action becomes the Einstein-frame sigma-model action
\eqref{eq:action_D}, with target metric
\begin{equation}\label{eq:einstein_target_from_jordan}
\G_{IJ}^{(E)}=\frac{\mathcal H_{IJ}}{\mathcal A}+\frac{D-1}{D-2}\partial_I\ln\mathcal A\,\partial_J\ln\mathcal A .
\end{equation}
Therefore an Einstein-frame Buchdahl--sigma solution gives the Jordan-frame solution
\begin{equation}\label{eq:jordan_solution_metric}
\dd\tilde s^2=\mathcal A(\Phi(\Sigma))^{-2/(D-2)}\dd s_\beta^2,
\quad
\varphi^I=\Phi^I(\Sigma),
\end{equation}
provided that $\Phi^I(\Sigma)$ is an affinely parametrized geodesic of $\G_{IJ}^{(E)}$ with speed $C_D(\beta)$,
\begin{equation}\label{eq:jordan_speed_general}
\G_{IJ}^{(E)}(\Phi)\frac{\dd\Phi^I}{\dd\Sigma}\frac{\dd\Phi^J}{\dd\Sigma}=C_D(\beta).
\end{equation}

The conformal factor changes areas, redshifts, curvature invariants and the matter-frame interpretation of singular surfaces. Null cones are unchanged where $\mathcal A$ is finite and positive, but physical circumferential radii and measured curvature scalars are Jordan-frame quantities.

We now give a genuinely two-scalar example. Let the Jordan-frame scalar coordinates be
\begin{equation}\label{eq:jordan_two_scalar_coordinates}
\varphi^I=(\phi,\eta),
\quad
I=1,2 ,
\end{equation}
and consider
\begin{align}\label{eq:jordan_two_scalar_action}
S_\J= \frac{1}{2\kappa_D}\int \dd^Dx\sqrt{|\tilde g|} &\left[
\mathcal A(\phi,\eta)\tilde R-\tilde g^{AB}\partial_A\phi\partial_B\phi \right.
\nonumber\\
&\left.-e^{2b\phi}\tilde g^{AB}\partial_A\eta\partial_B\eta
\right] .
\end{align}
Thus
\begin{equation}\label{eq:jordan_two_scalar_H}
\mathcal H_{IJ}\dd\varphi^I\dd\varphi^J
=
\dd\phi^2+e^{2b\phi}\dd\eta^2 .
\end{equation}

For definiteness, one can take the nonminimal coupling to be
\begin{equation}\label{eq:jordan_two_scalar_A}
\mathcal A(\phi,\eta)
=
e^{2\alpha\phi+2\delta\eta},
\end{equation}
with constants $\alpha$, $\delta$, and $b$. The Einstein-frame target
metric is then
\begin{eqnarray}\label{eq:jordan_two_scalar_E_target}
\dd\ell_E^2
&=&
\G_{IJ}^{(E)}\dd\varphi^I\dd\varphi^J
\nonumber\\
&=& e^{-2\alpha\phi-2\delta\eta} \left( \dd\phi^2+e^{2b\phi}\dd\eta^2 \right) \nonumber \\
&&+ \frac{4(D-1)}{D-2} \left( \alpha\dd\phi+\delta\dd\eta \right)^2 .
\end{eqnarray}
The Buchdahl construction now requires the two scalar fields to follow a
geodesic of the metric \eqref{eq:jordan_two_scalar_E_target}. Writing
\begin{equation}\label{eq:jordan_two_scalar_curve}
\phi=\phi(\Sigma),
\quad
\eta=\eta(\Sigma),
\end{equation}
the speed condition becomes
\begin{align}\label{eq:jordan_two_scalar_speed}
&
e^{-2\alpha\phi-2\delta\eta}
\left[
\left(\frac{\dd\phi}{\dd\Sigma}\right)^2
+
e^{2b\phi}
\left(\frac{\dd\eta}{\dd\Sigma}\right)^2
\right]
\nonumber\\
&\quad
+
\frac{4(D-1)}{D-2}
\left(
\alpha\frac{\dd\phi}{\dd\Sigma}
+
\delta\frac{\dd\eta}{\dd\Sigma}
\right)^2
=
C_D(\beta).
\end{align}
The curve also satisfies the two geodesic equations
\begin{equation}\label{eq:jordan_two_scalar_geodesic}
\frac{\dd^2\varphi^I}{\dd\Sigma^2}
+
\Gamma^{(E)I}{}_{JK}(\varphi)
\frac{\dd\varphi^J}{\dd\Sigma}
\frac{\dd\varphi^K}{\dd\Sigma}
=
0,
\quad
\varphi^I=(\phi,\eta),
\end{equation}
where $\Gamma^{(E)I}{}_{JK}$ is the Levi-Civita connection of
$\G_{IJ}^{(E)}$.

Given any Ricci-flat cyclic seed
\begin{equation}\label{eq:jordan_seed_metric}
\dd s_0^2
=
\eps e^{2\Sigma}\dd y^2
+
h_{ij}\dd x^i\dd x^j ,
\end{equation}
the corresponding Einstein-frame Buchdahl metric is
\begin{equation}\label{eq:jordan_einstein_buchdahl_metric}
\dd s_\beta^2
=
\eps e^{2\beta\Sigma}\dd y^2
+
e^{2(1-\beta)\Sigma/(D-3)}
h_{ij}\dd x^i\dd x^j .
\end{equation}
The full Jordan-frame two-scalar solution is therefore
\begin{eqnarray}\label{eq:jordan_two_scalar_full_solution}
\dd\tilde s^2&=&e^{-\frac{4}{D-2}[\alpha\phi(\Sigma)+\delta\eta(\Sigma)]} \left[\eps e^{2\beta\Sigma}\dd y^2\right.\nonumber\\
&&\left.+e^{2(1-\beta)\Sigma/(D-3)} h_{ij}\dd x^i\dd x^j \right],
\end{eqnarray}

Here $\phi(\Sigma)$ and $\eta(\Sigma)$ are determined by the target-space geodesic equations \eqref{eq:jordan_two_scalar_geodesic} and by the speed constraint \eqref{eq:jordan_two_scalar_speed}.

This example illustrates why the Jordan-frame extension is not merely a rewriting of the Einstein-frame result. In the Einstein frame, two target geodesics with the same speed source the same metric $g_{\beta AB}$. In the Jordan frame, however, the physical metric contains the position-dependent factor
\begin{equation}\label{eq:jordan_two_scalar_conformal_factor}
\mathcal A(\phi(\Sigma),\eta(\Sigma))^{-2/(D-2)} =e^{-\frac{4}{D-2}[\alpha\phi(\Sigma)+\delta\eta(\Sigma)]}.
\end{equation}

Thus two geodesics with the same speed but different trajectories in the
$(\phi,\eta)$ target space can lead to inequivalent Jordan-frame geometries.

A simple integrable subcase is obtained by setting
\begin{equation}\label{eq:jordan_two_scalar_integrable_subcase}
\alpha=\delta=0,
\quad
\mathcal A=1.
\end{equation}
Then the Jordan and Einstein frames coincide, but the target space remains nontrivial:
\begin{equation}\label{eq:jordan_two_scalar_hyperbolic_target}
\dd\ell_E^2=\dd\phi^2+e^{2b\phi}\dd\eta^2 .
\end{equation}
A nonradial geodesic of this hyperbolic target can be written as
\begin{align}\label{eq:jordan_two_scalar_hyperbolic_geodesic}
b\eta(\Sigma)&=X_c+R\tanh\!\left[b(s_0+\lambda\Sigma)\right],
\\
e^{-b\phi(\Sigma)}&=R\sech\!\left[b(s_0+\lambda\Sigma)\right],
\end{align}
with
\begin{equation}\label{eq:jordan_two_scalar_lambda}
\lambda^2=C_D(\beta).
\end{equation}

For nonzero $\alpha$ or $\delta$, the construction is unchanged in
principle, but the geodesics must be computed using the full Einstein-frame target metric \eqref{eq:jordan_two_scalar_E_target}. The conformal factor in Eq.~\eqref{eq:jordan_two_scalar_full_solution} then changes the physical Jordan-frame areas, redshifts and curvature invariants.

\section{Explicit target geometries, seed metrics and physical solutions}
\label{sec:families}

This section displays the solution generator in a form suitable for applications.
The input is not an arbitrary collection of scalar profiles. The input is a
Ricci-flat cyclic seed and the target metric $\G_{IJ}$ specified by the
matter theory. The scalar profiles are then obtained by solving the
target-space geodesic equations \eqref{eq:target_geodesic_D}, with affine
parameter $\Sigma$, subject to the fixed-speed condition
\eqref{eq:speed_condition_D}. The spacetime scalar fields are finally obtained
by the composition \eqref{eq:physical_curve_solution}.

The construction therefore separates two tasks. The spacetime task is universal:
given a seed of the form \eqref{eq:seed_metric_D}, the metric is generated by
Eq.~\eqref{eq:D_buch_metric}. The target-space task depends on the theory:
given $\G_{IJ}$, one computes geodesics $\Phi^I(\Sigma)$ with speed
$C_D(\beta)$. This section first lists physically motivated target metrics and
their corresponding geodesic profiles. It then inserts those profiles into
several Ricci-flat seeds.

We write
\begin{equation}\label{eq:lambda_def}
\lambda^2\equiv C_D(\beta)
=
\frac{D-2}{D-3}(1-\beta^2).
\end{equation}
For a positive-definite target metric, $\lambda$ is real only for
$|\beta|\leq1$. If $|\beta|>1$, the same spacetime metric can still be
supported by an indefinite target direction, a phantom scalar, or by complex
scalar fields. In what follows, unless otherwise stated, the target metric is
positive definite and $|\beta|\leq1$.

\subsection{Target-space geodesics from $\G_{IJ}$}

\subsubsection{Canonical moduli and nontrivial normalization matrix}

The simplest case is the flat target metric
\begin{equation}\label{eq:canonical_target_metric}
\dd\ell_\T^2
=
\G_{IJ}\dd\varphi^I\dd\varphi^J
=
\delta_{IJ}\dd\varphi^I\dd\varphi^J .
\end{equation}
The corresponding Einstein-frame scalar sector is
\begin{equation}\label{eq:canonical_target_action}
S_{\rm can}
=
\frac{1}{2\kappa_D}
\int \dd^Dx\sqrt{|g|}
\left[
R
-
\delta_{IJ}g^{AB}\partial_A\varphi^I\partial_B\varphi^J
\right].
\end{equation}

This describes $N$ canonically normalized massless scalars. It is the natural local form for weakly coupled moduli, for scalar sectors expanded around a regular point of target space, and for the standard multi-field generalization of the Fisher--Janis--Newman--Winicour--Wyman scalar solution. The geodesics are straight lines,
\begin{equation}\label{eq:canonical_target_solution}
\varphi^I(\Sigma)=\varphi^I_\infty+c^I\Sigma,
\quad
\delta_{IJ}c^Ic^J=\lambda^2 .
\end{equation}

The scalar charge is therefore a vector $c^I$ in target space. The metric depends only on its invariant norm, while the direction of $c^I$ determines which linear combination of fields is excited.

More generally, a set of free scalars may have a constant positive-definite kinetic matrix,
\begin{equation}\label{eq:constant_metric_target}
\dd\ell_\T^2= K_{IJ}\dd\varphi^I\dd\varphi^J,
\quad
K_{IJ}=\mathrm{constant},
\quad
K_{IJ}>0 .
\end{equation}
The corresponding scalar sector is
\begin{equation}\label{eq:constant_metric_target_action}
S_K=\frac{1}{2\kappa_D}\int \dd^Dx\sqrt{|g|}\left[R-K_{IJ}g^{AB}\partial_A\varphi^I\partial_B\varphi^J
\right].
\end{equation}

This includes fields with different normalizations and the local diagonalization of a general positive-definite target metric. The profiles are again linear,
\begin{equation}\label{eq:constant_metric_target_solution}
\varphi^I(\Sigma)=\varphi^I_\infty+c^I\Sigma,
\qquad
K_{IJ}c^Ic^J=\lambda^2 .
\end{equation}
Thus the invariant scalar charge is the $K$-norm of the charge vector. The individual components $c^I$ are coordinate-dependent, while $K_{IJ}c^Ic^J$ is the quantity fixed by the Buchdahl deformation.

\subsubsection{Compact spherical target}

Tensor--multiscalar theories and nonlinear sigma models often use curved target spaces, including maximally symmetric compact targets. A representative example is the two-sphere,
\begin{equation}\label{eq:spherical_target_metric}
\dd\ell_\T^2 = \ell^2 \left( \dd\Theta^2+\sin^2\Theta\,\dd\Phi_a^2 \right),
\end{equation}
where $\ell$ is the target-space radius. The corresponding Einstein-frame action is
\begin{align}\label{eq:spherical_target_action}
S_{S^2}=&\frac{1}{2\kappa_D}\int \dd^Dx\sqrt{|g|}\left[R-\ell^2 g^{AB}\partial_A\Theta\partial_B\Theta \right. \nonumber\\
&\left.-\ell^2\sin^2\Theta\,g^{AB}\partial_A\Phi_a\partial_B\Phi_a\right].
\end{align}

Thus the underlying theory contains two scalar fields. The Buchdahl construction does not reduce the theory to a one-field theory. Rather, it selects a one-dimensional geodesic subsector of the two-dimensional target. A simple representative geodesic is the meridian branch,
\begin{equation}\label{eq:spherical_target_meridian}
\Theta(\Sigma)
=
\Theta_0+\frac{\lambda}{\ell}\Sigma,
\quad
\Phi_a(\Sigma)=\Phi_{a0}.
\end{equation}
On this branch $\Phi_a$ is constant because the selected great circle is a
meridian. This is analogous to choosing a straight line along one coordinate
axis in a flat target. It is useful for calculations, but it does not display
the most general coordinate form of a spherical-target geodesic.

A great circle for which both angular coordinates vary can be written locally as
\begin{equation}\label{eq:spherical_target_general_u}
u(\Sigma)
=
u_0+\frac{\lambda}{\ell}\Sigma ,
\end{equation}
with
\begin{align}\label{eq:spherical_target_general_geodesic}
\cos\Theta(\Sigma)
&=
\sin\alpha\,\sin u(\Sigma),
\\
\tan\!\left[\Phi_a(\Sigma)-\Phi_{a0}\right]
&=
\cos\alpha\,\tan u(\Sigma).
\end{align}
For $0<\alpha<\pi/2$, both $\Theta$ and $\Phi_a$ vary in this chart. The
limits $\alpha=\pi/2$ and $\alpha=0$ give, respectively, a meridian-type
branch and an equatorial branch. Equation~\eqref{eq:spherical_target_general_geodesic}
is simply a great circle on $S^2$ written in spherical coordinates. It has
speed
\begin{equation}\label{eq:spherical_target_speed_check}
\ell^2\left[\left(\frac{\dd\Theta}{\dd\Sigma}\right)^2+\sin^2\Theta\left(\frac{\dd\Phi_a}{\dd\Sigma}\right)^2\right]=\lambda^2 .
\end{equation}

This coordinate expression may require different angular patches when the great circle crosses coordinate singularities, but the target-space curve itself is globally regular.

The Einstein-frame spacetime metric depends only on the invariant speed $\lambda^2$, not on the angular chart used to describe the great circle. Therefore the meridian branch and the two-coordinate branch can support the same Einstein-frame geometry, while corresponding to different coordinate descriptions of the same compact-target motion.

\subsubsection{Hyperbolic target}

A basic noncompact curved target is the upper half-plane,
\begin{equation}\label{eq:hyperbolic_metric}
\dd\ell_\T^2
=
\ell^2\frac{\dd X^2+\dd Y^2}{Y^2},
\quad
Y>0 .
\end{equation}
This is the Poincar\'e model of the two-dimensional hyperbolic space
$H^2$, equivalently the symmetric space
$SL(2,\mathbb R)/SO(2)$. It is a canonical negatively curved target for
harmonic-map and nonlinear-sigma-model systems, and it also appears naturally
in gravitational and string-theory reductions with noncompact duality groups
\cite{Helgason1978,Eells:1964,Misner:1978am,Breitenlohner:1987dg,Maharana:1992my}.
The corresponding Einstein-frame scalar sector is
\begin{align}\label{eq:hyperbolic_action}
S_{H^2}=&\frac{1}{2\kappa_D}\int \dd^Dx\sqrt{|g|}\left[R \right. \nonumber\\
&\left.-\ell^2\frac{g^{AB}\partial_A X\partial_B X+g^{AB}\partial_A Y\partial_B Y}{Y^2}\right].
\end{align}

This metric is the standard constant-negative-curvature geometry of $H^2$. It also appears in many moduli-space and axion--dilaton systems after a suitable field redefinition. A vertical geodesic gives
\begin{equation}\label{eq:hyperbolic_vertical}
X(\Sigma)=X_0,
\quad
Y(\Sigma)
=
Y_0\exp\!\left(\frac{\lambda}{\ell}\Sigma\right).
\end{equation}

A more general semicircular geodesic orthogonal to the boundary is
\begin{align}\label{eq:hyperbolic_semicircle}
X(\Sigma)&=X_c+R\tanh\!\left(u_0+\frac{\lambda}{\ell}\Sigma\right),
\\
Y(\Sigma)&=R\sech\!\left(u_0+\frac{\lambda}{\ell}\Sigma\right).
\end{align}

Both profiles have target speed $\lambda^2$. The vertical geodesic excites a single radial target direction, whereas the semicircle mixes the two target coordinates. Thus the same Buchdahl metric can represent a purely radial target-space flow or a mixed flow depending on the chosen target geodesic.

\subsubsection{Axion--dilaton target}

A particularly important configuration is the axion--dilaton metric
\begin{equation}\label{eq:axion_dilaton_metric}
\dd\ell_\T^2
=
\dd\phi^2+e^{2b\phi}\dd a^2 ,
\end{equation}
where $\phi$ is a dilaton, $a$ is an axion and $b$ controls the
normalization of the exponential coupling. This target geometry is the
two-scalar hyperbolic sector that appears in axion--dilaton gravity,
string-inspired black-hole systems and $SL(2,\mathbb R)/SO(2)$-type
duality models
\cite{Gibbons:1987ps,Garfinkle:1990qj,Shapere:1991ta,Sen:1992ua,Schwarz:1993vs,Maharana:1992my,Kallosh:1994ba}. The corresponding scalar sector is
\begin{align}\label{eq:axion_dilaton_action}
S_{\rm AD}=&\frac{1}{2\kappa_D}\int \dd^Dx\sqrt{|g|}
\left[R-g^{AB}\partial_A\phi\partial_B\phi \right. \\
&\left.-e^{2b\phi}g^{AB}\partial_Aa\partial_Ba
\right].
\end{align}
For $b\neq0$, the target metric \eqref{eq:axion_dilaton_metric} is equivalent to the upper-half-plane metric through
\begin{equation}
X=ba,
\quad
Y=e^{-b\phi},
\quad
\dd\ell_\T^2
=
\frac1{b^2}\frac{\dd X^2+\dd Y^2}{Y^2}.
\end{equation}
It is the scalar geometry underlying many string-inspired dilaton--axion
systems and $SL(2,\mathbb R)/SO(2)$-type cosets
\cite{Gibbons:1987ps,Kallosh:1994ba,Ortin:2004ms}. The special case $b=0$ is instead the flat two-scalar target and should be treated as the corresponding flat-target limit.

The vertical geodesic gives a pure dilaton profile,
\begin{equation}\label{eq:pure_dilaton_profile}
a(\Sigma)=a_0,
\quad
\phi(\Sigma)
=
\phi_0-\lambda\Sigma .
\end{equation}
A mixed axion--dilaton geodesic is
\begin{align}\label{eq:axion_dilaton_profiles}
ba(\Sigma)
&=
X_c+R\tanh\!\left(u_0+b\lambda\Sigma\right),
\\
e^{-b\phi(\Sigma)}
&=
R\sech\!\left(u_0+b\lambda\Sigma\right).
\end{align}
The invariant scalar charge is the target-space norm fixed by Eq.~\eqref{eq:lambda_def}. The separate axion and dilaton charges are coordinate-dependent components of the tangent vector to the same target geodesic.

\subsubsection{Jordan-frame targets}
For a Jordan-frame tensor--multiscalar theory, the relevant target metric for the Buchdahl construction is the Einstein-frame target metric \eqref{eq:einstein_target_from_jordan}. This is the standard target-space geometry of tensor--multiscalar gravity \cite{Damour:1992we}. Given $\mathcal A(\varphi)$ and $\mathcal H_{IJ}(\varphi)$, the scalar profiles are obtained by solving Eq.~\eqref{eq:target_geodesic_D} with $\G_{IJ}=\G^{(E)}_{IJ}$. The Jordan-frame metric is then reconstructed by Eq.~\eqref{eq:jordan_solution_metric}. In this case, two target geodesics with the same speed source the same Einstein-frame Buchdahl metric, but generally lead to inequivalent Jordan-frame geometries because the conformal factor depends on the position of the geodesic in target space.

\subsubsection{\texorpdfstring{Summary of the $\mathcal{G}_{IJ}$ target geometries}{Summary of the target geometries}}

In table~\ref{tab:universal_target_profiles} all the $\mathcal{G}_{IJ}$ considered in this paper are presented. The spacetime source is always proportional to $\partial_A\Sigma\,\partial_B\Sigma$, but the same source can be realized as a flat charge vector, a hyperbolic target-space geodesic, an axion--dilaton trajectory, or a compact-target great circle. This degeneracy is physical: in the Einstein frame these branches can share the same metric, while their scalar charges, target-space interpretation and Jordan-frame images differ.

\begin{table*}[t!]
\centering
\caption{Universal scalar profiles obtained by composing target-space geodesics
with the Buchdahl potential $\Sigma(x)$. The same Einstein-frame metric is
obtained for all profiles with the same invariant speed
$\lambda^2=C_D(\beta)$.}
\label{tab:universal_target_profiles}
\begin{ruledtabular}
\begin{tabular}{p{0.24\textwidth}p{0.50\textwidth}p{0.18\textwidth}}
Target realization & Scalar profile & Speed condition / comments \\
\hline
Canonical flat target
&
$\displaystyle
\varphi^I=\varphi_\infty^I+c^I\Sigma
$
&
$\displaystyle
\delta_{IJ}c^Ic^J=\lambda^2
$
\\[1.0em]

Constant $K_{IJ}$ target
&
$\displaystyle
\varphi^I=\varphi_\infty^I+c^I\Sigma
$
&
$\displaystyle
K_{IJ}c^Ic^J=\lambda^2
$
\\[1.0em]

Hyperbolic vertical branch
&
$\displaystyle
X=X_0,\quad
Y=Y_0\exp\!\left(\frac{\lambda}{\ell}\Sigma\right)
$
&
Geodesic on $H^2$
\\[1.0em]

Hyperbolic semicircle branch
&
$\displaystyle
X=X_c+R\tanh\!\left(u_0+\frac{\lambda}{\ell}\Sigma\right),
\quad
Y=R\sech\!\left(u_0+\frac{\lambda}{\ell}\Sigma\right)
$
&
Generic $H^2$ geodesic
\\[1.0em]

Pure axion--dilaton
&
$\displaystyle
a=a_0,\quad
\phi=\phi_0-\lambda\Sigma
$
&
Vertical hyperbolic geodesic
\\[1.0em]

Mixed axion--dilaton
&
$\displaystyle
ba=X_c+R\tanh\!\left(u_0+b\lambda\Sigma\right),
\quad
e^{-b\phi}=R\sech\!\left(u_0+b\lambda\Sigma\right)
$
&
Semicircular hyperbolic geodesic
\\[1.0em]

Spherical meridian
&
$\displaystyle
\Theta=\Theta_0+\frac{\lambda}{\ell}\Sigma,
\quad
\Phi_a=\Phi_{a0}
$
&
Great circle with constant azimuth
\\[1.0em]

Spherical two-coordinate branch
&
$\displaystyle
u=u_0+\frac{\lambda}{\ell}\Sigma,\quad
\cos\Theta=\sin\alpha\,\sin u,\quad
\tan(\Phi_a-\Phi_{a0})=\cos\alpha\,\tan u
$
&
Generic great circle on $S^2$
\end{tabular}
\end{ruledtabular}
\end{table*}

\subsection{Seed metrics and full solutions}

\subsubsection{General $D$-dimensional Tangherlini seed}

A classical solution to the $D$-dimensional vacuum Einstein field equations is the Schwarzschild--Tangherlini seed given by~\cite{Tangherlini:1963bw}
\begin{equation}\label{eq:tangherlini_seed}
\dd s_0^2=-F\dd t^2+F^{-1}\dd r^2+r^2\dd\Omega_{D-2}^2,
\end{equation}
where
\begin{equation}
F(r)=1-\left(\frac{r_0}{r}\right)^{D-3}.
\end{equation}

Considering $y=t$, $\eps=-1$ we get
\begin{equation}\label{eq:tangherlini_sigma}
\Sigma(r)=\frac12\ln F(r).
\end{equation}
Equation~\eqref{eq:D_buch_metric} gives the $D$-dimensional scalar-dressed
Tangherlini family
\begin{align}\label{eq:tangherlini_generated}
\dd s_\beta^2
&=
-F^\beta\dd t^2
+
F^{(1-\beta)/(D-3)}
\left[
F^{-1}\dd r^2+r^2\dd\Omega_{D-2}^2
\right],
\\
\varphi^I(r)&=\Phi^I\!\left(\frac12\ln F\right),
\quad
\G_{IJ}(\Phi)\Phi^{I\prime}\Phi^{J\prime}=\frac{D-2}{D-3}(1-\beta^2).
\end{align}

For the canonical target,
\begin{equation}\label{eq:tangherlini_canonical_scalars}
\varphi^I(r)=\varphi_\infty^I+\frac{c^I}{2}\ln F,
\quad
\delta_{IJ}c^Ic^J=\frac{D-2}{D-3}(1-\beta^2).
\end{equation}
For a constant normalized target $K_{IJ}$,
\begin{equation}\label{eq:tangherlini_K_scalars}
\varphi^I(r)=\varphi_\infty^I+\frac{c^I}{2}\ln F,
\quad
K_{IJ}c^Ic^J=\frac{D-2}{D-3}(1-\beta^2).
\end{equation}
For the hyperbolic target \eqref{eq:hyperbolic_metric}, the vertical branch is
\begin{equation}\label{eq:tangherlini_hyperbolic_vertical}
X(r)=X_0,
\quad
Y(r)
=
Y_0F^{\lambda/(2\ell)} ,
\end{equation}
and the semicircular branch is
\begin{align}\label{eq:tangherlini_hyperbolic_semicircle}
X(r)
&=
X_c+R\tanh\!\left[
u_0+\frac{\lambda}{2\ell}\ln F
\right],
\\
Y(r)
&=
R\sech\!\left[
u_0+\frac{\lambda}{2\ell}\ln F
\right].
\end{align}

For the axion--dilaton target \eqref{eq:axion_dilaton_metric}, the pure
dilaton branch is
\begin{equation}\label{eq:tangherlini_pure_dilaton}
a(r)=a_0,
\quad
\phi(r)
=
\phi_0-\frac{\lambda}{2}\ln F,
\end{equation}
whereas the mixed axion--dilaton branch is
\begin{align}\label{eq:tangherlini_axion_dilaton}
ba(r)
&=
X_c+R\tanh\!\left[
u_0+\frac{b\lambda}{2}\ln F
\right],
\\
e^{-b\phi(r)}
&=
R\sech\!\left[
u_0+\frac{b\lambda}{2}\ln F
\right].
\end{align}
For the compact spherical target \eqref{eq:spherical_target_metric}, the
meridian branch gives
\begin{equation}\label{eq:tangherlini_spherical_target_scalars}
\Theta(r)
=
\Theta_0+\frac{\lambda}{2\ell}\ln F,
\quad
\Phi_a(r)=\Phi_{a0}.
\end{equation}
The two-coordinate great-circle branch is
\begin{align}\label{eq:tangherlini_spherical_general_scalars}
u(r)
&=
u_0+\frac{\lambda}{2\ell}\ln F,
\\
\cos\Theta(r)
&=
\sin\alpha\,\sin u(r),
\\
\tan\!\left[\Phi_a(r)-\Phi_{a0}\right]
&=
\cos\alpha\,\tan u(r).
\end{align}

The different branches
\eqref{eq:tangherlini_canonical_scalars}--\eqref{eq:tangherlini_spherical_general_scalars}
describe the same Einstein-frame metric whenever their target speeds agree.
What changes is the matter interpretation. The canonical branch carries a
constant charge vector in a flat target. The $K_{IJ}$ branch shows explicitly
that only the target-space norm of this vector is fixed by geometry. The
hyperbolic branch realizes the same source as geodesic motion on a negatively
curved target. The axion--dilaton branch is a physically important
parametrization of the same hyperbolic geometry, where the invariant charge is
distributed nonlinearly between axionic and dilatonic components. The compact
branch realizes the same source as motion along a great circle of $S^2$,
either in a meridian chart or in a chart where both angular fields vary. Thus
the Buchdahl parameter fixes the total scalar kinetic strength, while the target
geometry determines how that strength is represented by physical fields.

The Ricci scalar is fixed by the universal identity
\begin{equation}\label{eq:tangherlini_ricci_scalar}
R[g_\beta]
=
C_D(\beta)g_\beta^{rr}(\partial_r\Sigma)^2,
\quad
\partial_r\Sigma
=
\frac{D-3}{2}\frac{r_0^{D-3}}{r^{D-2}F}.
\end{equation}
For a positive-definite target and nonzero scalar speed, the surface $F=0$ is
not a regular event horizon. The scalar kinetic invariant diverges there. Thus
the Buchdahl deformation converts the Tangherlini black-hole horizon into a
curvature singularity, in agreement with the usual no-hair expectation for
minimally coupled massless scalars~\cite{Agnese:1985xj}.

The area radius of the symmetry spheres is
\begin{equation}\label{eq:tangherlini_areal_radius}
R_{\rm sph}(r)
=
r\,F^{(1-\beta)/[2(D-3)]}.
\end{equation}
For the nontrivial positive-definite-target branch $-1<\beta<1$, $R_{\rm sph}\to0$ as $r\to r_0$, even though the coordinate value $r=r_0$ is finite. If one additionally restricts to positive ADM mass, this becomes $0<\beta<1$. Using Eq.~\eqref{eq:tangherlini_ricci_scalar}, the curvature
divergence can be written explicitly as
\begin{equation}\label{eq:tangherlini_ricci_near_singularity}
R[g_\beta]
=
\frac{(D-2)(D-3)}{4}(1-\beta^2)
\frac{r_0^{2(D-3)}}{r^{2(D-2)}}
F^{-1-\frac{1-\beta}{D-3}} .
\end{equation}
Therefore, for every nonzero positive-target scalar speed, $-1<\beta<1$, the surface $r=r_0$ is a genuine curvature singularity rather than a regular Killing horizon. The Schwarzschild--Tangherlini black-hole seed is recovered only at $\beta=1$. The other zero-speed endpoint, $\beta=-1$, has a constant positive-target scalar map but gives the reciprocal Ricci-flat member rather than the original Tangherlini black hole.

A useful invariant way of seeing the nature of the surface $F=0$ is to compute the radial proper distance to it. Near $r=r_0$,
\begin{equation}\label{eq:phys_tangherlini_proper_distance}
\ell_{\rm prop}
\sim
\int F^{\frac{1-\beta}{2(D-3)}-\frac12}\,\dd r
\sim
F^{\frac12+\frac{1-\beta}{2(D-3)}} .
\end{equation}
For the positive-target branch $|\beta|\leq1$, this distance is finite. Thus
the scalar singularity is not pushed to an asymptotic end of the geometry: it
is reached at finite spatial distance, while the area radius
\eqref{eq:tangherlini_areal_radius} collapses for $-1\leq\beta<1$.

At spatial infinity $F\to1$, the geometry is asymptotically flat and the
fields approach constant target-space values. For the canonical branch,
\begin{equation}\label{eq:tangherlini_asymptotic_scalars}
\varphi^I(r)
=
\varphi_\infty^I
-
\frac{c^I}{2}
\left(\frac{r_0}{r}\right)^{D-3}
+
O\!\left(r^{-2(D-3)}\right).
\end{equation}

Moreover, the ADM mass is rescaled by the Buchdahl parameter,
\begin{equation}\label{eq:tangherlini_mass_D}
M_\beta = \beta M_0,
\quad
M_0 = \frac{(D-2)\Omega_{D-2}}{16\pi G_D}r_0^{D-3},
\end{equation}
and the invariant scalar charge is controlled by
\begin{equation}\label{eq:tangherlini_charge_norm_D}
q^I=-\frac12 c^I r_0^{D-3},
\quad
\delta_{IJ}q^Iq^J
=
\frac14 r_0^{2(D-3)}
\frac{D-2}{D-3}(1-\beta^2).
\end{equation}
This makes the trade-off between gravitational mass and scalar charge explicit:
for fixed seed scale $r_0$, decreasing $\beta$ transfers part of the
Tangherlini redshift potential into scalar kinetic energy.

It is useful and important to make the physical meaning of the scalar charge and the status of the would-be horizon more explicit. For the canonical target, define the invariantly normalized asymptotic scalar-charge magnitude
\begin{equation}\label{eq:tangherlini_normalized_scalar_charge}
\mathcal Q_{\rm s}^{\,2}
\equiv
\frac{(D-2)(D-3)\Omega_{D-2}^{\,2}}
     {64\pi^2G_D^{\,2}}\,
\delta_{IJ}q^Iq^J .
\end{equation}
Equations~\eqref{eq:tangherlini_mass_D} and
\eqref{eq:tangherlini_charge_norm_D} then imply
\begin{equation}\label{eq:tangherlini_mass_charge_circle}
M_\beta^2+\mathcal Q_{\rm s}^{\,2}=M_0^2 .
\end{equation}
Thus, at fixed seed scale $r_0$, the Buchdahl parameter redistributes the invariant scale between ADM mass and asymptotic scalar charge. The scalar charge measures the departure from the vacuum Schwarzschild--Tangherlini endpoint; it is not, on the nontrivial positive-target branch, a regular black-hole charge supported by a horizon.

The absence of a regular horizon can also be seen directly from the radial null geometry. Defining
\begin{equation}\label{eq:tangherlini_a_def}
a\equiv\frac{1-\beta}{D-3},
\end{equation}
radial null curves of Eq.~\eqref{eq:tangherlini_generated} satisfy
\begin{equation}\label{eq:tangherlini_radial_null}
\frac{\dd t}{\dd r}
=
\pm F^{\frac{a-1-\beta}{2}} .
\end{equation}
Consequently, a tortoise-type coordinate obeys, close to the simple
zero $r=r_0$,
\begin{align}\label{eq:tangherlini_tortoise_near_r0}
r_*-r_{*,0}
&=
\int_{r_0}^{r}
F^{\frac{a-1-\beta}{2}}\,\dd r
\nonumber\\
&\propto
F^{\frac{(D-2)(1-\beta)}{2(D-3)}} ,
\qquad
-1<\beta<1 .
\end{align}
Hence $r_*$ remains finite as $r\to r_0$ throughout the nontrivial positive-target branch. This is qualitatively different from the logarithmic divergence of the Schwarzschild--Tangherlini tortoise coordinate at $\beta=1$. Together with Eq.~\eqref{eq:tangherlini_ricci_near_singularity}, this shows that $r=r_0$ is a singular boundary and not a regular event horizon.

For the positive-mass branch $0<\beta<1$, the same conclusion follows from the formal static surface-gravity expression. At the candidate surface one finds
\begin{align}\label{eq:tangherlini_formal_kappa}
\kappa_{\rm cand}
&=
\frac{1}{2}
\frac{\partial_r F^\beta}
{\sqrt{F^\beta F^{a-1}}}
\nonumber\\
&=
\frac{\beta F'}{2}
F^{-\frac{(D-2)(1-\beta)}{2(D-3)}} .
\end{align}
This expression diverges as $F\to0$ for $0<\beta<1$. It should
not be interpreted as a Hawking temperature, because the limiting
surface is curvature singular and is not a regular Killing horizon.

Consequently, the nontrivial positive-target members of the
Tangherlini-derived family do not possess a regular black-hole horizon
area, Hawking temperature, Bekenstein--Hawking entropy, or a horizon
scalar potential entering a black-hole first law. The only regular
Schwarzschild--Tangherlini black hole in this positive-target family
is the seed endpoint $\beta=1$. At that endpoint,
\begin{equation}\label{eq:tangherlini_seed_thermodynamics}
r_H=r_0,
\qquad
T_H=\frac{D-3}{4\pi r_0},
\qquad
S_H=\frac{\Omega_{D-2}r_0^{D-2}}{4G_D},
\qquad
q^I=0 ,
\end{equation}
and the standard first law reduces to
\begin{equation}\label{eq:tangherlini_seed_first_law}
\dd M_0=T_H\,\dd S_H .
\end{equation}
Thus, with the asymptotic target-space point held fixed, there is no additional scalar work term at the regular black-hole endpoint.

The relevant $D=4$ case gives 
\begin{equation}
F(r)=1-\frac{2m}{r},
\end{equation}
and Eq.~\eqref{eq:tangherlini_generated} reduces to the multi-field Fisher--Janis--Newman--Winicour--Wyman geometry
\begin{equation}\label{eq:JNW_multi_metric}
\dd s_\beta^2
=
-F^\beta\dd t^2
+
F^{-\beta}\dd r^2
+
F^{1-\beta}r^2\dd\Omega^2 .
\end{equation}
For the canonical target,
\begin{equation}\label{eq:JNW_canonical_scalars}
\varphi^I(r)
=
\varphi_\infty^I+\frac{c^I}{2}\ln F,
\quad
\delta_{IJ}c^Ic^J=2(1-\beta^2).
\end{equation}
For a constant target metric $K_{IJ}$,
\begin{equation}\label{eq:JNW_K_scalars}
\varphi^I(r)
=
\varphi_\infty^I+\frac{c^I}{2}\ln F,
\quad
K_{IJ}c^Ic^J=2(1-\beta^2).
\end{equation}
For the hyperbolic target \eqref{eq:hyperbolic_metric}, the vertical branch is
\begin{equation}\label{eq:JNW_hyperbolic_vertical}
X(r)=X_0,
\quad
Y(r)
=
Y_0 F^{\sqrt{2(1-\beta^2)}/(2\ell)} ,
\end{equation}
and the semicircular branch is
\begin{align}\label{eq:JNW_hyperbolic_semicircle}
X(r)
&=
X_c+R\tanh\!\left[
u_0+\frac{\sqrt{2(1-\beta^2)}}{2\ell}\ln F
\right],
\\
Y(r)
&=
R\sech\!\left[
u_0+\frac{\sqrt{2(1-\beta^2)}}{2\ell}\ln F
\right].
\end{align}

Writing the four-dimensional ADM mass in geometric units as
\begin{equation}
\mathcal M\equiv G_4 M_{\rm ADM},
\end{equation}
the mass and scalar charges in the canonical branch obey
\begin{equation}\label{eq:JNW_mass_charge_relation}
\mathcal M=\beta m,
\qquad
Q^I=-mc^I,
\qquad
\mathcal M^2+\frac12\delta_{IJ}Q^IQ^J=m^2 .
\end{equation}
Thus $m$ is the invariant radius scale, while $\beta$ controls the split between the geometric ADM mass and the scalar charge.

For the axion--dilaton target, the same metric \eqref{eq:JNW_multi_metric} is
supported either by the pure dilaton branch
\begin{equation}\label{eq:JNW_pure_dilaton}
a(r)=a_0,
\quad
\phi(r)
=
\phi_0-\frac{\sqrt{2(1-\beta^2)}}{2}\ln F,
\end{equation}
or by the mixed axion--dilaton branch
\begin{align}\label{eq:JNW_axion_dilaton}
ba(r)
&=
X_c+R\tanh\!\left[
u_0+\frac{b\sqrt{2(1-\beta^2)}}{2}\ln F
\right],
\\
e^{-b\phi(r)}
&=
R\sech\!\left[
u_0+\frac{b\sqrt{2(1-\beta^2)}}{2}\ln F
\right].
\end{align}
For the compact spherical target, the four-dimensional meridian branch is
\begin{equation}\label{eq:JNW_spherical_meridian}
\Theta(r)
=
\Theta_0+\frac{\sqrt{2(1-\beta^2)}}{2\ell}\ln F,
\quad
\Phi_a(r)=\Phi_{a0},
\end{equation}
and the two-coordinate branch is
\begin{align}\label{eq:JNW_spherical_general}
u(r)
&=
u_0+\frac{\sqrt{2(1-\beta^2)}}{2\ell}\ln F,
\\
\cos\Theta(r)
&=
\sin\alpha\,\sin u(r),
\\
\tan\!\left[\Phi_a(r)-\Phi_{a0}\right]
&=
\cos\alpha\,\tan u(r).
\end{align}

This illustrates an important physical point: the Einstein-frame geometry
depends only on the target-space speed, not on the coordinate components of the
target-space motion. A flat charge vector, a hyperbolic geodesic, a purely
dilatonic flow, a mixed axion--dilaton flow and a compact great-circle flow can
therefore source the same metric, while carrying different scalar
interpretations.

The circular photon orbit of \eqref{eq:JNW_multi_metric}, when it lies outside
the singular surface, is located at
\begin{equation}\label{eq:photon_sphere}
r_{\rm ph}=m(1+2\beta).
\end{equation}
It is outside $r=2m$ only for $\beta>1/2$. Hence the scalar charge does not
only remove the Schwarzschild horizon; it also changes the optical structure of
the exterior geometry.

\subsubsection{Neutral black strings and black branes}

A direct product of a Tangherlini black hole with $p$ flat directions gives a Ricci-flat neutral black-brane seed in total dimension $D$, with $p\geq0$ and $D-p-3>0$,
\cite{Tangherlini:1963bw,Horowitz:1991cd,Gregory:1993vy,Emparan:2008eg}
\begin{equation}\label{eq:black_brane_seed}
\dd s_0^2=-F\dd t^2+\sum_{\alpha=1}^{p}\dd z_\alpha^2+F^{-1}\dd r^2+r^2\dd\Omega_{D-p-2}^2,
\end{equation}
where
\begin{equation}
F(r)=1-\left(\frac{r_0}{r}\right)^{D-p-3}.
\end{equation}
The Buchdahl coordinate is again $y=t$, so
\begin{equation}
\Sigma(r)=\frac12\ln F(r).
\end{equation}

The generated metric solution is
\begin{eqnarray}\label{eq:black_brane_generated}
\dd s_\beta^2
&=&-F^\beta\dd t^2+F^{(1-\beta)/(D-3)}\times \nonumber \\
&&\left[\sum_{\alpha=1}^{p}\dd z_\alpha^2+F^{-1}\dd r^2+
r^2\dd\Omega_{D-p-2}^2
\right],
\end{eqnarray}
and the scalar fields have the form
\begin{align}
\varphi^I(r)
&=
\Phi^I\!\left(\frac12\ln F\right),
\quad
\G_{IJ}\Phi^{I\prime}\Phi^{J\prime}=\frac{D-2}{D-3}(1-\beta^2).
\end{align}

For canonical scalars one has
\begin{equation}\label{eq:black_brane_canonical_scalars}
\varphi^I(r)=\varphi_\infty^I+\frac{c^I}{2}\ln F,
\quad
\delta_{IJ}c^Ic^J=\frac{D-2}{D-3}(1-\beta^2),
\end{equation}
and for a constant target metric $K_{IJ}$,
\begin{equation}\label{eq:black_brane_K_scalars}
\varphi^I(r)=\varphi_\infty^I+\frac{c^I}{2}\ln F,
\quad
K_{IJ}c^Ic^J=\frac{D-2}{D-3}(1-\beta^2).
\end{equation}

For the hyperbolic target \eqref{eq:hyperbolic_metric}, the vertical branch is
\begin{equation}\label{eq:black_brane_hyperbolic_vertical}
X(r)=X_0,
\quad
Y(r)=Y_0F^{\lambda/(2\ell)} ,
\end{equation}
and the semicircular branch is
\begin{align}\label{eq:black_brane_hyperbolic_semicircle}
X(r)&=X_c+R\tanh\!\left[u_0+\frac{\lambda}{2\ell}\ln F\right],
\\
Y(r)&=R\sech\!\left[u_0+\frac{\lambda}{2\ell}\ln F\right].
\end{align}

For the axion--dilaton target,
\begin{equation}\label{eq:black_brane_pure_dilaton}
a(r)=a_0,
\quad
\phi(r)=\phi_0-\frac{\lambda}{2}\ln F,
\end{equation}
and
\begin{align}\label{eq:black_brane_axion_dilaton}
ba(r)
&=
X_c+R\tanh\!\left[
u_0+\frac{b\lambda}{2}\ln F
\right],
\\
e^{-b\phi(r)}
&=
R\sech\!\left[
u_0+\frac{b\lambda}{2}\ln F
\right].
\end{align}
For the compact spherical target, the meridian branch is
\begin{equation}\label{eq:black_brane_spherical_meridian}
\Theta(r)=\Theta_0+\frac{\lambda}{2\ell}\ln F,
\quad
\Phi_a(r)=\Phi_{a0},
\end{equation}
while the two-coordinate great-circle branch is
\begin{align}\label{eq:black_brane_spherical_general}
u(r)&=u_0+\frac{\lambda}{2\ell}\ln F,
\\
\cos\Theta(r)&=\sin\alpha\,\sin u(r),
\\
\tan\!\left[\Phi_a(r)-\Phi_{a0}\right]&=\cos\alpha\,\tan u(r).
\end{align}

The exponent in the Buchdahl conformal factor contains the total spacetime dimension $D$, not the transverse black-hole dimension $D-p$. This is because the cyclic density identity is applied to the full $D$-dimensional geometry, including the $p$ flat worldvolume directions. Consequently the brane directions $z_\alpha$ are not spectators after the deformation: they inherit the same conformal dressing as the transverse spatial block. This creates scalar-dressed descendants of neutral black-brane seeds whose worldvolume metric is no longer a direct product with the transverse geometry. For nonzero scalar speed, the seed horizon is generically replaced by a singular surface, as in the spherical Tangherlini case. The new feature relative to the spherical solution is that the scalar dressing also changes the effective scaling of the extended worldvolume directions. Hyperbolic, axion--dilaton and compact-target branches give inequivalent scalar interpretations of the same Einstein-frame brane geometry.

The asymptotic expansion makes this worldvolume effect explicit. Writing
\begin{equation}\label{eq:phys_brane_mu_def}
n=D-p-3,
\qquad
\mu=r_0^n,
\qquad
F=1-\frac{\mu}{r^n},
\end{equation}
one finds, for large $r$,
\begin{align}\label{eq:phys_brane_asymptotics}
g_{tt}
&=-1+\beta\frac{\mu}{r^n}+O(r^{-2n}),
\\
g_{z_\alpha z_\alpha}&=1-\frac{1-\beta}{D-3}\frac{\mu}{r^n}+O(r^{-2n}),
\\
\varphi^I&=\varphi_\infty^I-\frac{c^I}{2}\frac{\mu}{r^n}+O(r^{-2n})
\end{align}
for the canonical branch. Hence the same Buchdahl parameter that changes the asymptotic coefficient in $g_{tt}$ also induces scalar charge density and modifies the asymptotic coefficients along the extended brane directions. In an asymptotically Kaluza--Klein interpretation, the physical ADM mass density and gravitational tensions are determined by the appropriate combinations of these coefficients. We therefore refrain here from identifying the $g_{tt}$ coefficient alone with the full mass density; the robust conclusion is that the scalar dressing changes both the transverse black-object asymptotics and the worldvolume tension data.

\subsubsection{Static Weyl seeds and scalar multipoles in $D=4$}
Higher-dimensional static vacuum geometries with several commuting orthogonal Killing directions are naturally described by the generalized Weyl class~\cite{Emparan:2001wk}. These solutions provide further Ricci-flat cyclic seeds for the present construction whenever one isolates a non-null hypersurface-orthogonal cyclic block. In the following we focus on the four-dimensional Weyl sector, where the seed potential and its scalar descendants can be written explicitly. Every four-dimensional static axisymmetric vacuum metric can be written locally in Weyl form,
\begin{eqnarray}\label{eq:weyl_seed}
\dd s_0^2 &=&-e^{2U(\rho,z)}\dd t^2+e^{-2U(\rho,z)}
\times \nonumber \\
&&\left[e^{2k(\rho,z)}(\dd\rho^2+\dd z^2)+\rho^2\dd\phi^2\right].
\end{eqnarray}
This is of the seed form \eqref{eq:seed_metric_D} with $y=t$, $\eps=-1$, and
\begin{equation}\label{eq:weyl_sigma_definition}
\Sigma(\rho,z)
=
U(\rho,z)
=
\frac12\ln |g^{(0)}_{tt}| .
\end{equation}
The vacuum Einstein equations imply
\begin{equation}\label{eq:weyl_laplace}
\partial_\rho^2U+\frac1\rho\partial_\rho U+\partial_z^2U=0,
\end{equation}
with $k(\rho,z)$ determined by the usual first-order Weyl equations. Hence
any static Weyl vacuum geometry is a valid Ricci-flat, hypersurface-orthogonal
cyclic seed.

Since $D=4$, the generated metric is
\begin{equation}\label{eq:weyl_generated}
\dd s_\beta^2
=
-e^{2\beta U}\dd t^2
+
e^{-2\beta U}
\left[
e^{2k}(\dd\rho^2+\dd z^2)+\rho^2\dd\phi^2
\right],
\end{equation}
with
\begin{equation}
\varphi^I=\Phi^I(U),
\end{equation}
For canonical scalars,
\begin{equation}\label{eq:weyl_canonical_scalars}
\varphi^I(\rho,z)
=
\varphi_\infty^I+c^I U(\rho,z),
\quad
\delta_{IJ}c^Ic^J=2(1-\beta^2).
\end{equation}
For a constant target metric $K_{IJ}$,
\begin{equation}\label{eq:weyl_K_scalars}
\varphi^I(\rho,z)
=
\varphi_\infty^I+c^I U(\rho,z),
\quad
K_{IJ}c^Ic^J=2(1-\beta^2).
\end{equation}
For the hyperbolic target \eqref{eq:hyperbolic_metric}, the vertical branch is
\begin{equation}\label{eq:weyl_hyperbolic_vertical}
X(\rho,z)=X_0,
\quad
Y(\rho,z)
=
Y_0\exp\!\left[
\frac{\sqrt{2(1-\beta^2)}}{\ell}U(\rho,z)
\right],
\end{equation}
and the semicircular branch is
\begin{align}\label{eq:weyl_hyperbolic_semicircle}
X(\rho,z)
&=
X_c+R\tanh\!\left[
u_0+\frac{\sqrt{2(1-\beta^2)}}{\ell}U(\rho,z)
\right],
\\
Y(\rho,z)
&=
R\sech\!\left[
u_0+\frac{\sqrt{2(1-\beta^2)}}{\ell}U(\rho,z)
\right].
\end{align}
For axion--dilaton scalars,
\begin{equation}\label{eq:weyl_pure_dilaton}
a(\rho,z)=a_0,
\quad
\phi(\rho,z)
=
\phi_0-\sqrt{2(1-\beta^2)}\,U(\rho,z),
\end{equation}
and
\begin{align}\label{eq:weyl_axion_dilaton_scalars}
ba(\rho,z)
&=
X_c+R\tanh\!\left[
u_0+b\sqrt{2(1-\beta^2)}\,U(\rho,z)
\right],
\\
e^{-b\phi(\rho,z)}
&=
R\sech\!\left[
u_0+b\sqrt{2(1-\beta^2)}\,U(\rho,z)
\right].
\end{align}
For the compact spherical target, the meridian branch is
\begin{equation}\label{eq:weyl_spherical_meridian}
\Theta(\rho,z)
=
\Theta_0+\frac{\sqrt{2(1-\beta^2)}}{\ell}U(\rho,z),
\quad
\Phi_a(\rho,z)=\Phi_{a0},
\end{equation}
and the two-coordinate branch is
\begin{align}\label{eq:weyl_spherical_general}
u(\rho,z)
&=
u_0+\frac{\sqrt{2(1-\beta^2)}}{\ell}U(\rho,z),
\\
\cos\Theta(\rho,z)
&=
\sin\alpha\,\sin u(\rho,z),
\\
\tan\!\left[\Phi_a(\rho,z)-\Phi_{a0}\right]
&=
\cos\alpha\,\tan u(\rho,z).
\end{align}

The function $k$ is unchanged by the Buchdahl deformation. Hence conical defects, struts or rods present in the Weyl seed are inherited by the scalar-dressed solution. The Newtonian multipoles encoded in $U$ are rescaled in $g_{tt}$ by $\beta$, while the generated scalar multipoles are locked to the same harmonic potential. Canonical Weyl-type solutions with multipolar massless scalar fields have been studied recently in Ref.~\cite{Lim:2026cky}; accordingly, the novelty of the present subsection is not the existence of scalar Weyl multipoles by itself. Rather, the unified Buchdahl theorem gives the particularly constrained scalarization rule $\varphi^I=\Phi^I(U)$ and promotes the same seed multipole potential to an arbitrary target-space geodesic without solving a new coupled Einstein--sigma boundary-value problem. The scalar map inherits the rod, strut and multipolar structure of the seed through the single potential $U(\rho,z)$, while curved targets make its internal scalar realization intrinsically nonlinear even though the spacetime generating potential remains harmonic.

For an asymptotically flat Weyl seed with expansion
\begin{equation}\label{eq:phys_weyl_multipole_expansion}
U(R,\theta)=-\sum_{\ell=0}^{\infty}\frac{M_\ell}{R^{\ell+1}}P_\ell(\cos\theta),
\end{equation}
the weak-field gravitational multipoles appearing in $g_{tt}$ are rescaled by $\beta$, while the canonical scalar multipoles are
\begin{equation}\label{eq:phys_weyl_scalar_multipoles}
Q_\ell^I=-c^I M_\ell .
\end{equation}

Thus the generator gives an explicit multipolar scalarization rule: every Weyl multipole of the seed is copied into the sigma-model charge distribution along one target-space geodesic.

For the Curzon--Chazy potential,
\begin{equation}\label{eq:curzon_potential}
U(\rho,z)=-\frac{m}{\sqrt{\rho^2+z^2}},
\end{equation}
Eqs.~\eqref{eq:weyl_generated} and \eqref{eq:weyl_canonical_scalars} give the canonical scalar Curzon family
\begin{equation}\label{eq:curzon_canonical_scalar}
\varphi^I(\rho,z)=\varphi_\infty^I-\frac{m c^I}{\sqrt{\rho^2+z^2}} .
\end{equation}
The directional singularity of the Curzon seed is not removed; it is dressed by a target-space scalar charge distribution aligned with the same harmonic potential.

For a Zipoy--Voorhees seed,
\begin{equation}\label{eq:zv_weyl_data}
U_{\rm ZV}=\delta\,U_{\rm Schw}^{\rm Weyl},
\quad
k_{\rm ZV}=\delta^2 k_{\rm Schw}^{\rm Weyl}.
\end{equation}

The generated metric is Eq.~\eqref{eq:weyl_generated} with $U=U_{\rm ZV}$ and $k=k_{\rm ZV}$. The parameter $\delta$ controls the vacuum quadrupolar deformation of the seed, whereas $\beta$ controls the transfer of the cyclic potential into the scalar sector. The resulting scalar charge carries the same monopole and higher multipoles as $U_{\rm ZV}$, with target-space direction fixed by $\Phi^{I\prime}$.

\subsubsection{Kasner cosmologies}

The construction is not restricted to static metrics. Consider a $D$-dimensional Kasner seed of the form~\cite{Kasner:1921zz}
\begin{equation}\label{eq:kasner_seed}
\dd s_0^2 = -\dd t^2 + \sum_{a=1}^{D-1}t^{2p_a}(\dd x^a)^2,
\end{equation}
with
\begin{equation}\label{eq:kasner_conditions}
\sum_{a=1}^{D-1}p_a=1,
\quad
\sum_{a=1}^{D-1}p_a^2=1 .
\end{equation}
Choose one spatial cyclic coordinate $y=x^q$. Then
\begin{equation}
\eps=+1,
\quad
\Sigma(t)=p_q\ln t.
\end{equation}
The Buchdahl--sigma cosmology is given by
\begin{align}\label{eq:kasner_generated}
\dd s_\beta^2&=t^{2\beta p_q}(\dd x^q)^2+t^{2(1-\beta)p_q/(D-3)} \times \\
&\left[-\dd t^2+\sum_{a\neq q}t^{2p_a}(\dd x^a)^2\right],
\end{align}
with the scalars
\begin{align}
\varphi^I(t)
&=
\Phi^I(p_q\ln t),
\quad
\G_{IJ}\Phi^{I\prime}\Phi^{J\prime}=\frac{D-2}{D-3}(1-\beta^2).
\end{align}
For canonical scalars,
\begin{equation}\label{eq:kasner_canonical_scalars}
\varphi^I(t)=\varphi_\infty^I+p_q c^I\ln t,
\quad
\delta_{IJ}c^Ic^J=\frac{D-2}{D-3}(1-\beta^2).
\end{equation}
For a constant target metric $K_{IJ}$,
\begin{equation}\label{eq:kasner_K_scalars}
\varphi^I(t)=\varphi_\infty^I+p_q c^I\ln t,
\quad
K_{IJ}c^Ic^J=\frac{D-2}{D-3}(1-\beta^2).
\end{equation}
For the hyperbolic target \eqref{eq:hyperbolic_metric}, the vertical branch is
\begin{equation}\label{eq:kasner_hyperbolic_vertical}
X(t)=X_0,
\quad
Y(t)=Y_0t^{\lambda p_q/\ell},
\end{equation}
and the semicircular branch is
\begin{align}\label{eq:kasner_hyperbolic_semicircle}
X(t)&=X_c+R\tanh\!\left[u_0+\frac{\lambda p_q}{\ell}\ln t\right],
\\
Y(t)&=R\sech\!\left[u_0+\frac{\lambda p_q}{\ell}\ln t\right].
\end{align}

For the axion--dilaton target one has
\begin{equation}\label{eq:kasner_pure_dilaton}
a(t)=a_0,
\quad
\phi(t)=\phi_0-\lambda p_q\ln t,
\end{equation}
and
\begin{align}\label{eq:kasner_axion_dilaton}
ba(t)&=X_c+R\tanh\!\left[u_0+b\lambda p_q\ln t\right],
\\
e^{-b\phi(t)}&=R\sech\!\left[u_0+b\lambda p_q\ln t\right].
\end{align}

For the compact spherical target, the meridian branch is given by
\begin{equation}\label{eq:kasner_spherical_meridian}
\Theta(t)=\Theta_0+\frac{\lambda p_q}{\ell}\ln t,
\quad
\Phi_a(t)=\Phi_{a0},
\end{equation}
and the two-coordinate branch is
\begin{align}\label{eq:kasner_spherical_general}
u(t)&=u_0+\frac{\lambda p_q}{\ell}\ln t,
\\
\cos\Theta(t)&=\sin\alpha\,\sin u(t),
\\
\tan\!\left[\Phi_a(t)-\Phi_{a0}\right]&=\cos\alpha\,\tan u(t).
\end{align}

Moreover, it is useful to rewrite the metric in cosmic time. Defining
\begin{equation}\label{eq:phys_kasner_A_def}
A=\frac{(1-\beta)p_q}{D-3},
\qquad
\dd T=t^A\dd t ,
\end{equation}
for $A\neq -1$, the generated geometry is again of Kasner type in the Einstein frame, being given by
\begin{equation}\label{eq:phys_kasner_cosmic_metric}
\dd s_\beta^2=-\dd T^2+T^{2\tilde p_q}(\dd x^q)^2+\sum_{a\neq q}T^{2\tilde p_a}(\dd x^a)^2,
\end{equation}
with scalar-dressed exponents
\begin{equation}\label{eq:phys_kasner_exponents}
\tilde p_q=\frac{\beta p_q}{1+A},
\qquad
\tilde p_a=\frac{p_a+A}{1+A}
\quad 
(a\neq q).
\end{equation}

Particularizing for the canonical scalars one has
\begin{equation}\label{eq:phys_kasner_scalar_charges}
\varphi^I(T)=\varphi_0^I+q^I\ln T,
\qquad
q^I=\frac{p_q c^I}{1+A},
\end{equation}
with the exponents obeying the scalar-Kasner relations
\begin{equation}\label{eq:phys_kasner_relations}
\sum_{a=1}^{D-1}\tilde p_a=1,
\qquad
\sum_{a=1}^{D-1}\tilde p_a^2+\delta_{IJ}q^Iq^J=1 ,
\end{equation}

The Buchdahl map therefore moves a vacuum Kasner point on the Kasner sphere to a scalar-Kasner point inside it, with the displacement controlled by the selected cyclic exponent $p_q$ and the target speed. Furthermore, although the Buchdahl direction is spacelike, the scalar gradient is timelike because $\Sigma=p_q\ln t$. These solutions are therefore anisotropic scalar cosmologies generated from vacuum Kasner metrics. The new physics is that a single vacuum Kasner exponent $p_q$ controls both the conformal dressing of the remaining directions and the affine speed with which the scalar map moves through target space. In a flat target this gives logarithmic moduli evolution. In the hyperbolic target it gives cosmological geodesic motion on $H^2$. In the axion--dilaton target it gives a string-inspired realization of the same hyperbolic flow. In the compact target it gives repeated great-circle motion as the universe approaches or recedes from the Kasner singularity.

\subsubsection{Buchdahl--sigma Kaluza--Klein bubbles}

A particularly interesting spacelike-cyclic seed is the Lorentzian product of time with the Euclidean Tangherlini instanton. This is a static Kaluza--Klein Killing-bubble geometry, closely related to the geometry underlying Witten's bubble-of-nothing construction but not identical to Witten's time-dependent expanding bubble~\cite{Witten:1981gj,Kastor:2008wd}. In dimension $D\geq5$, consider
\begin{equation}\label{eq:kk_bubble_seed}
\dd s_0^2=-\dd t^2+F(r)\dd\psi^2+F(r)^{-1}\dd r^2+r^2\dd\Omega_{D-3}^2,
\end{equation}
with
\begin{equation}
F(r)=1-\left(\frac{r_0}{r}\right)^{D-4}.
\end{equation}

The cyclic coordinate considered is the spacelike Kaluza--Klein circle,
\begin{equation}\label{eq:kk_bubble_sigma}
y=\psi,
\quad
\eps=+1,
\quad
\Sigma(r)=\frac12\ln F(r).
\end{equation}
Equation~\eqref{eq:D_buch_metric} gives
\begin{eqnarray}\label{eq:kk_bubble_generated}
\dd s_\beta^2
&=&F^\beta \dd\psi^2+F^{(1-\beta)/(D-3)}\times \nonumber \\
&&\left[-\dd t^2+F^{-1}\dd r^2+r^2\dd\Omega_{D-3}^2\right],
\end{eqnarray}
and the scalar-fields have the form
\begin{equation}
\varphi^I(r)=\Phi^I\!\left(\frac12\ln F\right),\quad
\G_{IJ}(\Phi)\Phi^{I\prime}\Phi^{J\prime}=\frac{D-2}{D-3}(1-\beta^2).
\end{equation}

For the canonical target,
\begin{equation}\label{eq:kk_bubble_canonical}
\varphi^I(r)
=
\varphi_\infty^I+\frac{c^I}{2}\ln F,
\quad
\delta_{IJ}c^Ic^J
=
\frac{D-2}{D-3}(1-\beta^2).
\end{equation}
For a constant target metric $K_{IJ}$,
\begin{equation}\label{eq:kk_bubble_K}
\varphi^I(r)
=
\varphi_\infty^I+\frac{c^I}{2}\ln F,
\quad
K_{IJ}c^Ic^J
=
\frac{D-2}{D-3}(1-\beta^2).
\end{equation}
For the hyperbolic target \eqref{eq:hyperbolic_metric}, the vertical branch is
\begin{equation}\label{eq:kk_bubble_hyperbolic_vertical}
X(r)=X_0,
\quad
Y(r)=Y_0F^{\lambda/(2\ell)} ,
\end{equation}
and the semicircular branch is
\begin{align}\label{eq:kk_bubble_hyperbolic_semicircle}
X(r)&=X_c+R\tanh\!\left[u_0+\frac{\lambda}{2\ell}\ln F\right],
\\
Y(r)&=R\sech\!\left[u_0+\frac{\lambda}{2\ell}\ln F\right].
\end{align}

For the axion--dilaton target \eqref{eq:axion_dilaton_metric}, the pure dilaton branch is
\begin{equation}\label{eq:kk_bubble_pure_dilaton}
a(r)=a_0,
\quad
\phi(r)=\phi_0-\frac{\lambda}{2}\ln F,
\end{equation}
and the mixed axion--dilaton branch is
\begin{align}\label{eq:kk_bubble_axion_dilaton}
ba(r)&=X_c+R\tanh\!\left[u_0+\frac{b\lambda}{2}\ln F\right],
\\
e^{-b\phi(r)}&=R\sech\!\left[u_0+\frac{b\lambda}{2}\ln F\right].
\end{align}

For the compact spherical target \eqref{eq:spherical_target_metric}, the meridian branch is
\begin{equation}\label{eq:kk_bubble_spherical_meridian}
\Theta(r)=\Theta_0+\frac{\lambda}{2\ell}\ln F,
\quad
\Phi_a(r)=\Phi_{a0},
\end{equation}
while the two-coordinate great-circle branch is
\begin{align}\label{eq:kk_bubble_spherical_general}
u(r)&=u_0+\frac{\lambda}{2\ell}\ln F,
\\
\cos\Theta(r)&=\sin\alpha\,\sin u(r),
\\
\tan\!\left[\Phi_a(r)-\Phi_{a0}\right]&=\cos\alpha\,\tan u(r).
\end{align}

This is a scalar-dressed Kaluza--Klein bubble generated from a spacelike cyclic norm rather than from a redshift factor. The seed bubble can be made regular at $r=r_0$ by choosing the period of $\psi$ appropriately. For the nontrivial positive-target branch $-1<\beta<1$, however, the scalar speed is nonzero and the scalar kinetic invariant typically diverges at $F=0$. At $\beta=-1$ the positive-target scalar map is again constant, but the metric is the reciprocal Ricci-flat member rather than the smooth seed bubble. This can also be seen from the local bubble geometry. Since
\begin{equation}\label{eq:phys_bubble_local_powers}
g_{\psi\psi}\sim F^\beta,
\qquad
g_{rr}\sim F^{-1+\frac{1-\beta}{D-3}},
\qquad
F\sim r-r_0 ,
\end{equation}
the proper radial distance behaves as
\begin{equation}\label{eq:phys_bubble_rho}
\rho\sim(r-r_0)^{\frac12\left[1+\frac{1-\beta}{D-3}\right]},
\end{equation}
whereas the circumference of the Kaluza--Klein circle behaves as
\begin{equation}\label{eq:phys_bubble_circumference}
C_\psi\sim(r-r_0)^{\beta/2}.
\end{equation}
The curvature singularity can also be displayed explicitly. Since
\begin{equation}
\partial_r\Sigma=\frac{D-4}{2} \frac{r_0^{D-4}}{r^{D-3}F},
\end{equation}
the universal Ricci-source identity gives
\begin{equation}\label{eq:kk_bubble_ricci_exact}
R[g_\beta]= \frac{(D-2)(D-4)^2}{4(D-3)} (1-\beta^2) \frac{r_0^{2(D-4)}}{r^{2(D-3)}} F^{-1-\frac{1-\beta}{D-3}} .
\end{equation}
Thus $r=r_0$ is a genuine curvature singularity for every nontrivial positive-target deformation.

It is also not a black-hole horizon. Radial null curves at fixed $\psi$ and fixed angles satisfy
\begin{equation}\label{eq:kk_bubble_radial_null}
\frac{\dd t}{\dd r}=\pm F^{-1/2},
\end{equation}
so the corresponding null-coordinate distance to $r=r_0$ is finite. The vanishing of $g_{tt}$ produced by the common Buchdahl conformal factor therefore does not define a regular Killing horizon.

For completeness, at the undeformed endpoint $\beta=1$ the $(r,\psi)$ geometry is smooth if the Kaluza--Klein coordinate has period
\begin{equation}\label{eq:kk_bubble_seed_period}
\Delta\psi
=
\frac{4\pi}{F'(r_0)}
=
\frac{4\pi r_0}{D-4}.
\end{equation}
For $\beta\neq1$, however, the mismatch between the powers in Eqs.~\eqref{eq:phys_bubble_rho} and \eqref{eq:phys_bubble_circumference} cannot be removed by any choice of $\Delta\psi$.

A smooth origin would require $C_\psi\propto\rho$, which gives $\beta=1$. Thus the Buchdahl deformation converts the smooth bubble surface into a scalar singularity for every nontrivial positive-target deformation. This is the bubble analogue of the way the JNW deformation converts the Schwarzschild horizon into a curvature singularity. The family is therefore not a regular bubble of nothing for generic $\beta$, but it is an exact arbitrary-target scalar bubble geometry and a useful spacelike-cyclic counterpart of the Tangherlini construction. It is also distinct from recent Kaluza--Klein bubble constructions with different matter sectors, such as massive scalar fields~\cite{Jackson:2025bpn}.

The canonical Kaluza--Klein bubble branch also carries an asymptotic
scalar charge. Expanding Eq.~\eqref{eq:kk_bubble_canonical} at large
$r$ gives
\begin{equation}\label{eq:kk_bubble_asymptotic_charge}
\varphi^I(r)=\varphi_\infty^I-\frac{c^I}{2}\left(\frac{r_0}{r}\right)^{D-4}+O\!\left(r^{-2(D-4)}\right),
\end{equation}
so that it is natural to define
\begin{equation}\label{eq:kk_bubble_scalar_charge}
q_{\rm KK}^I=-\frac12 c^I r_0^{D-4}.
\end{equation}
This is an asymptotic Kaluza--Klein scalar charge rather than a black-hole horizon charge. For the nontrivial positive-target branch the geometry has no regular horizon and hence no associated black-hole temperature, entropy, or horizon scalar potential. At $\beta=1$ the scalar charge vanishes and one recovers the smooth vacuum Kaluza--Klein bubble, whose relevant global characteristics are those of an asymptotically Kaluza--Klein spacetime rather than black-hole thermodynamic quantities.

\paragraph{Alternative azimuthal cyclic directions.}
The spherical factors of the Tangherlini and Kaluza--Klein bubble seeds provide additional spacelike cyclic directions. Buchdahl transformations with respect to spacelike, and in particular azimuthal, Killing directions have been considered previously in the four-dimensional vacuum setting; for example, applying the reciprocal transformation to Schwarzschild with respect to $\partial_\phi$ leads to a Schwarzschild--Levi-Civita-type geometry \cite{Akbar:2015qna,Barrientos:2024uuq}. Here the same possibility follows directly from the general cyclicconstruction developed above.

Let $\phi$ denote the final azimuthal coordinate on a unit $N$-sphere,
so that
\begin{equation}\label{eq:angular_seed_norm}
g_{\phi\phi}^{(0)}
=r^2\mathcal P_N^2,
\quad
\mathcal P_N=\prod_{a=1}^{N-1}\sin\theta_a ,
\end{equation}
where $N=D-2$ for the Tangherlini seed and $N=D-3$ for the Kaluza--Klein bubble seed.

On every connected patch excluding the rotation axes
$\mathcal P_N=0$, the vector $\partial_\phi$ is a non-nullhypersurface-orthogonal cyclic direction and may therefore be used in the Buchdahl construction. The associated potential is
\begin{equation}\label{eq:angular_buchdahl_potential}
\Sigma_\phi=\frac12\ln g_{\phi\phi}^{(0)}=\ln(r\mathcal P_N).
\end{equation}
The corresponding generated metric takes the compact form
\begin{equation}\label{eq:angular_buchdahl_generated}
\dd s_{\beta,\phi}^2
=\left(r^2\mathcal P_N^2\right)^\beta\dd\phi^2+\left(r^2\mathcal P_N^2\right)^{\frac{1-\beta}{D-3}}
\left[\dd s_0^2-r^2\mathcal P_N^2\dd\phi^2
\right],
\end{equation}
while the scalar fields are
\begin{equation}\label{eq:angular_buchdahl_scalars}
\varphi^I=\Phi^I(\Sigma_\phi),
\quad
\mathcal G_{IJ}(\Phi)\frac{\dd\Phi^I}{\dd\Sigma_\phi}\frac{\dd\Phi^J}{\dd\Sigma_\phi}=C_D(\beta).
\end{equation}

Thus the azimuthal Killing direction does generate an additional family of exact solutions locally. There are, however, important global qualifications. First, $g_{\phi\phi}^{(0)}$ vanishes on the rotation axes, where
\begin{equation}
\Sigma_\phi\longrightarrow-\infty .
\end{equation}
The construction must therefore be understood patchwise, and a nontrivial positive-target scalar profile is generically singular on the axes. This is analogous to the axial singular behavior encountered in four-dimensional azimuthal Buchdahl transformations of the Schwarzschild geometry \cite{Barrientos:2024uuq}. Second,
\begin{equation}
\Sigma_\phi\sim\ln r
\quad
(r\to\infty),
\end{equation}
so the nontrivial scalar does not approach a constant at infinity and the generated geometry does not preserve the standard asymptotically flat or asymptotically Kaluza--Klein behavior of the corresponding seed. The appearance of a Levi-Civita-type asymptotic sector in the four-dimensional vacuum analogue is likewise familiar from Refs.~\cite{Akbar:2015qna,Barrientos:2024uuq}. The angularly generated branches considered here are therefore valid local exact solutions, but they do not furnish globally regular scalar-hairy versions of the original Tangherlini black hole or Kaluza--Klein bubble.

\subsubsection{Null-gradient Rosen-wave Buchdahl--sigma solutions}

The solution generator also produces non-static wave solutions. Consider the four-dimensional Ricci-flat Rosen seed
\begin{equation}\label{eq:rosen_seed}
\dd s_0^2=-2\dd u\dd v+a(u)^2\dd y^2+b(u)^2\dd z^2,
\end{equation}
with
\begin{equation}\label{eq:rosen_vacuum_condition}
\frac{a''(u)}{a(u)}+\frac{b''(u)}{b(u)}=0.
\end{equation}

This is the standard Rosen form of a plane-wave geometry, closely related to the Baldwin--Jeffery, Rosen and Bondi--Pirani--Robinson descriptions of exact plane waves~\cite{Baldwin:1926,rosen1937plane,Bondi:1958aj,Griffiths:1991zp}. The cyclic coordinate is spacelike, hence
\begin{equation}\label{eq:rosen_sigma}
y=y,
\quad
\eps=+1,
\quad
\Sigma(u)=\ln a(u).
\end{equation}

Since $D=4$, Eq.~\eqref{eq:D_buch_metric} gives
\begin{align}\label{eq:rosen_generated}
\dd s_\beta^2&=a(u)^{2\beta}\dd y^2+a(u)^{2(1-\beta)}\left[-2\dd u\dd v+b(u)^2\dd z^2\right],
\\
\varphi^I(u)&=\Phi^I(\ln a(u)),
\quad
\G_{IJ}(\Phi)\Phi^{I\prime}\Phi^{J\prime}=2(1-\beta^2).
\end{align}

For canonical scalars the scalar fields have the form
\begin{equation}\label{eq:rosen_canonical}
\varphi^I(u)=\varphi_0^I+c^I\ln a(u),
\quad
\delta_{IJ}c^Ic^J=2(1-\beta^2),
\end{equation}
and for a constant target metric $K_{IJ}$,
\begin{equation}\label{eq:rosen_K}
\varphi^I(u)=\varphi_0^I+c^I\ln a(u),
\quad
K_{IJ}c^Ic^J=2(1-\beta^2).
\end{equation}

For the hyperbolic target \eqref{eq:hyperbolic_metric}, the vertical branch is
\begin{equation}\label{eq:rosen_hyperbolic_vertical}
X(u)=X_0,
\quad
Y(u)=Y_0a(u)^{\sqrt{2(1-\beta^2)}/\ell},
\end{equation}
and the semicircular branch is
\begin{align}\label{eq:rosen_hyperbolic_semicircle}
X(u)&=X_c+R\tanh\!\left[u_0+\frac{\sqrt{2(1-\beta^2)}}{\ell}\ln a(u)
\right],
\\
Y(u)&=R\sech\!\left[u_0+\frac{\sqrt{2(1-\beta^2)}}{\ell}\ln a(u)
\right].
\end{align}

For the axion--dilaton target,
\begin{equation}\label{eq:rosen_pure_dilaton}
a_{\rm ax}(u)=a_0,
\quad
\phi(u)=\phi_0-\sqrt{2(1-\beta^2)}\,\ln a(u),
\end{equation}
and
\begin{align}\label{eq:rosen_axion_dilaton}
b\,a_{\rm ax}(u)&=X_c+R\tanh\!\left[u_0+b\sqrt{2(1-\beta^2)}\,\ln a(u)\right],
\\
e^{-b\phi(u)}&=R\sech\!\left[u_0+b\sqrt{2(1-\beta^2)}\,\ln a(u)\right].
\end{align}
Here $a_{\rm ax}$ denotes the axion field, to avoid confusion with the Rosen scale factor $a(u)$. For the compact spherical target, the meridian branch is
\begin{equation}\label{eq:rosen_spherical_meridian}
\Theta(u)=\Theta_0+\frac{\sqrt{2(1-\beta^2)}}{\ell}\ln a(u),
\quad
\Phi_a(u)=\Phi_{a0},
\end{equation}
and the two-coordinate branch is
\begin{align}\label{eq:rosen_spherical_general}
s(u)&=s_0+\frac{\sqrt{2(1-\beta^2)}}{\ell}\ln a(u),
\\
\cos\Theta(u)&=\sin\alpha\,\sin s(u),
\\
\tan\!\left[\Phi_a(u)-\Phi_{a0}\right]&=\cos\alpha\,\tan s(u).
\end{align}

These solutions are qualitatively different from the static Buchdahl
families and provide a direct example of the null-gradient sector
identified above. Since $\Sigma=\ln a(u)$ depends only on the null
coordinate $u$, one has $\dd\Sigma\neq0$ wherever $a'(u)\neq0$, while
\begin{equation}\label{eq:rosen_null_gradient}
(\nabla\Sigma)^2=0 .
\end{equation}
The scalar curvature therefore vanishes,
\begin{equation}\label{eq:phys_rosen_scalar_curvature}
R[g_\beta]=C_D(\beta)(\nabla\Sigma)^2=0,
\end{equation}
but the Ricci tensor carries a nonzero null component,
\begin{equation}\label{eq:phys_rosen_null_ricci}
R_{uu}[g_\beta]=2(1-\beta^2)\left(\frac{\dd}{\dd u}\ln a(u)\right)^2 .
\end{equation}
Equivalently, the sigma-model stress tensor is of pure radiation type,
\begin{equation}\label{eq:phys_rosen_null_flux}
\G_{IJ}\frac{\dd\varphi^I}{\dd u}\frac{\dd\varphi^J}{\dd u}=2(1-\beta^2)\left(\frac{\dd}{\dd u}\ln a(u)\right)^2 .
\end{equation}

The Buchdahl parameter controls the amplitude of a nonlinear-sigma null wave rather than the strength of a radial scalar charge. The profiles displayed above belong to the universal target-geodesic subclass, although the null-gradient property permits the more general constant-speed target curves noted after Eq.~\eqref{eq:main_scalar_field_check}. Thus the Rosen family gives exact Einstein--sigma wave solutions from any Ricci-flat Rosen seed satisfying Eq.~\eqref{eq:rosen_vacuum_condition}. It is naturally compared with recent parallel-wave solutions in Einstein--nonlinear-sigma models~\cite{Bonga:2022nwe}.

\subsubsection{Polarized Gowdy Buchdahl--sigma cosmologies}

The same mechanism applies to inhomogeneous cosmological seeds. Consider the standard polarized Gowdy $T^3$ vacuum metric~\cite{Gowdy:1971jh,Gowdy:1973mu}
\begin{equation}\label{eq:gowdy_seed}
\dd s_0^2 = e^{\Lambda/2} \left( -e^{-2\tau}\dd\tau^2+\dd\theta^2 \right) + e^{-\tau} \left( e^P\dd\sigma^2+e^{-P}\dd\delta^2 \right),
\end{equation}
where $P=P(\tau,\theta)$ satisfies the polarized Gowdy wave equation
\begin{equation}\label{eq:gowdy_wave}
P_{\tau\tau}-e^{-2\tau}P_{\theta\theta}=0 .
\end{equation}

The function $\Lambda=\Lambda(\tau,\theta)$ is determined by the constraints
\begin{equation}\label{eq:gowdy_constraints_Lambda}
\Lambda_\tau=1-P_\tau^2-e^{-2\tau}P_\theta^2,
\quad
\Lambda_\theta=-2P_\tau P_\theta .
\end{equation}
Equivalently, if the usual Gowdy function $\lambda_G$ is used, then
$\Lambda=\tau-\lambda_G$, and the seed metric can be written as
\begin{eqnarray}\label{eq:gowdy_seed_lambdaG}
\dd s_0^2&=&e^{(\tau-\lambda_G)/2}
\left(-e^{-2\tau}\dd\tau^2+\dd\theta^2
\right)\nonumber\\
&&+e^{-\tau}
\left(e^P\dd\sigma^2+e^{-P}\dd\delta^2\right),
\end{eqnarray}
with
\begin{equation}\label{eq:gowdy_constraints_lambdaG}
(\lambda_G)_\tau=P_\tau^2+e^{-2\tau}P_\theta^2,
\quad
(\lambda_G)_\theta=2P_\tau P_\theta .
\end{equation}
This seed belongs to the class of compact vacuum Gowdy cosmologies with two commuting spacelike Killing fields~\cite{Gowdy:1971jh,Gowdy:1973mu,Ringstrom:2010zz,Ringstrom:2009}.

Choosing the spacelike cyclic coordinate
\begin{equation}\label{eq:gowdy_sigma}
y=\sigma,\;
\eps=+1, \;
e^{2\Sigma}=e^{P-\tau}, \;
\Sigma(\tau,\theta)=\frac12\left[P(\tau,\theta)-\tau\right],
\end{equation}
for $D=4$, Eq.~\eqref{eq:D_buch_metric} gives the generated metric
\begin{eqnarray}\label{eq:gowdy_generated}
\dd s_\beta^2&=&e^{\beta(P-\tau)}\dd\sigma^2+e^{(1-\beta)(P-\tau)} \times \nonumber \\
&&\left[e^{\Lambda/2}\left(-e^{-2\tau}\dd\tau^2+\dd\theta^2
\right)+e^{-\tau-P}\dd\delta^2\right],
\end{eqnarray}
and the scalar fields are given by
\begin{align}\label{eq:gowdy_scalar_general}
\varphi^I(\tau,\theta)&=\Phi^I\!\left(
\frac12[P(\tau,\theta)-\tau]\right).
\end{align}

For canonical scalars one has
\begin{equation}\label{eq:gowdy_canonical}
\varphi^I(\tau,\theta)=\varphi_0^I+\frac{c^I}{2}\left[P(\tau,\theta)-\tau\right],
\quad
\delta_{IJ}c^Ic^J=2(1-\beta^2),
\end{equation}
and similarly, for a constant target metric $K_{IJ}$,
\begin{equation}\label{eq:gowdy_K}
\varphi^I(\tau,\theta)=\varphi_0^I + \frac{c^I}{2}\left[P(\tau,\theta)-\tau\right],
\quad
K_{IJ}c^Ic^J=2(1-\beta^2).
\end{equation}
For the hyperbolic target,
\begin{align}\label{eq:gowdy_hyperbolic_vertical}
X(\tau,\theta)&=X_0,
\\
Y(\tau,\theta)&=Y_0\exp\!\left[
\frac{\sqrt{2(1-\beta^2)}}{2\ell}(P-\tau)
\right].
\end{align}
and the semicircular branch is
\begin{align}\label{eq:gowdy_hyperbolic_semicircle}
X(\tau,\theta)&=X_c+R\tanh\!\left[u_0+\frac{\sqrt{2(1-\beta^2)}}{2\ell}\left(P-\tau\right)\right],
\\
Y(\tau,\theta)&=R\sech\!\left[u_0+\frac{\sqrt{2(1-\beta^2)}}{2\ell}\left(P-\tau\right)\right].
\end{align}

For the axion--dilaton target, let $a_{\rm ax}$ denote the axion field. The pure dilaton branch is
\begin{equation}\label{eq:gowdy_pure_dilaton}
a_{\rm ax}(\tau,\theta)=a_0,
\quad
\phi(\tau,\theta)
=
\phi_0
-
\frac{\sqrt{2(1-\beta^2)}}{2}
\left(P-\tau\right),
\end{equation}
whereas the mixed axion--dilaton branch is
\begin{align}\label{eq:gowdy_axion_dilaton}
b\,a_{\rm ax}(\tau,\theta)
&=
X_c+R\tanh\!\left[
u_0+
\frac{b\sqrt{2(1-\beta^2)}}{2}
\left(P-\tau\right)
\right],
\\
e^{-b\phi(\tau,\theta)}
&=
R\sech\!\left[
u_0+
\frac{b\sqrt{2(1-\beta^2)}}{2}
\left(P-\tau\right)
\right].
\end{align}

For the compact spherical target, the meridian branch is
\begin{equation}\label{eq:gowdy_spherical_meridian}
\Theta(\tau,\theta)
=
\Theta_0+
\frac{\sqrt{2(1-\beta^2)}}{2\ell}
\left(P-\tau\right),
\quad
\Phi_a(\tau,\theta)=\Phi_{a0},
\end{equation}
and the two-coordinate great-circle branch is
\begin{align}\label{eq:gowdy_spherical_general}
s(\tau,\theta)
&=
s_0+
\frac{\sqrt{2(1-\beta^2)}}{2\ell}
\left(P-\tau\right),
\\
\cos\Theta(\tau,\theta)
&=
\sin\alpha\,\sin s(\tau,\theta),
\\
\tan\!\left[\Phi_a(\tau,\theta)-\Phi_{a0}\right]
&=
\cos\alpha\,\tan s(\tau,\theta).
\end{align}

This family is an inhomogeneous cosmological Buchdahl--sigma solution. The Buchdahl potential is not radial and not purely time-dependent; it is the Gowdy combination
\begin{equation}\label{eq:gowdy_buchdahl_potential}
\Sigma=\frac12(P-\tau).
\end{equation}
Hence the scalar fields inherit both the time dependence and the spatial inhomogeneity of the polarized Gowdy wave. The result is an exact nonlinear sigma-model cosmology in which the full scalar map is controlled by one vacuum Gowdy potential and one target-space geodesic.

Near a velocity-dominated Gowdy singularity one often writes locally
\begin{equation}\label{eq:phys_gowdy_velocity_ansatz}
P(\tau,\theta)
\sim
v(\theta)\tau+P_0(\theta),
\qquad
\tau\to+\infty .
\end{equation}
Then the Buchdahl potential behaves as
\begin{equation}\label{eq:phys_gowdy_sigma_velocity}
\Sigma(\tau,\theta)
\sim
\frac12\left[v(\theta)-1\right]\tau+\frac12P_0(\theta).
\end{equation}

For any positive-definite target, the leading scalar kinetic density along the target geodesic is therefore controlled by
\begin{equation}\label{eq:phys_gowdy_scalar_velocity}
\G_{IJ}\partial_\tau\varphi^I\partial_\tau\varphi^J
\sim
\frac14 C_4(\beta)\left[v(\theta)-1\right]^2 .
\end{equation}
Thus the leading temporal contribution to the target-space kinetic density vanishes at points where the Gowdy velocity equals the areal contraction rate, $v(\theta)=1$, and is strongest where the polarized wave amplitude deviates most from that rate. This statement refers to the temporal component displayed in Eq.~\eqref{eq:phys_gowdy_scalar_velocity}. The full Buchdahl--sigma source is
\begin{equation}\label{eq:phys_gowdy_full_source}
R_{AB}[g_\beta] = C_4(\beta)\,\partial_A\Sigma\,\partial_B\Sigma .
\end{equation}
Therefore the full scalar source vanishes only where the full gradient
$\partial_A\Sigma$ vanishes. Indeed, the velocity-dominated expansion gives
\begin{equation}\label{eq:phys_gowdy_theta_gradient}
\partial_\theta\Sigma \sim \frac12 v'(\theta)\tau+\frac12 P_0'(\theta),
\end{equation}
so angular source components may remain nonzero even at points where $v(\theta)=1$. The generated scalar energy is therefore controlled by the full local gradient of the Gowdy Buchdahl potential, with Eq.~\eqref{eq:phys_gowdy_scalar_velocity} describing its leading temporal part.

\subsubsection{Accelerating target-geodesic C-metric family}

As a final static but non-spherical example, consider the uncharged vacuum C-metric seed. The C-metric describes accelerating black holes and has several standard equivalent forms~\cite{KinnersleyWalker1970,HongTeo2003,GriffithsKrtousPodolsky2006,GriffithsPodolsky2009}. With the conformal factor chosen as $\Omega=A(x-\eta)$, a Ricci-flat form is
\begin{equation}\label{eq:cmetric_seed}
\dd s_0^2=\frac{1}{\Omega^2}\left[-F(\eta)\dd t^2+\frac{\dd\eta^2}{F(\eta)}+\frac{\dd x^2}{G(x)}+G(x)\dd\phi^2\right],
\end{equation}
where
\begin{equation}\label{eq:cmetric_FG}
G(x)=1-x^2-2mAx^3,
\quad
F(\eta)=-1+\eta^2+2mA\eta^3 .
\end{equation}

The plus sign in the cubic term of $F(\eta)$ is tied to the choice
$\Omega=A(x-\eta)$. Equivalently, one may use the convention
$\Omega=A(x+\eta)$, in which case the corresponding function is
$F(\eta)=-1+\eta^2-2mA\eta^3$. Choosing the temporal cyclic coordinate
\begin{equation}\label{eq:cmetric_sigma}
y=t,\;
\eps=-1, \;
e^{2\Sigma}=\frac{F(\eta)}{\Omega^2}, \;
\Sigma(x,\eta)
=
\frac12\ln\!\left(\frac{F(\eta)}{\Omega^2}\right).
\end{equation}
This choice is made in a static patch where $F(\eta)>0$, $G(x)>0$, and $\Omega\neq0$. The formulas below are restricted to this connected fixed-sign domain. For $D=4$, Eq.~\eqref{eq:D_buch_metric} gives
\begin{eqnarray}\label{eq:cmetric_generated}
\dd s_\beta^2&=&-\left(\frac{F}{\Omega^2}\right)^\beta\dd t^2+\left(\frac{F}{\Omega^2}\right)^{1-\beta}\frac{1}{\Omega^2} \times \nonumber\\ 
&&\left[\frac{\dd\eta^2}{F}+\frac{\dd x^2}{G}+G\dd\phi^2\right],
\end{eqnarray}
and
\begin{equation}
\varphi^I(x,\eta)=\Phi^I\!\left[\frac12\ln\!\left(\frac{F(\eta)}{\Omega^2}\right)\right].
\end{equation}

For canonical scalars,
\begin{equation}\label{eq:cmetric_canonical}
\varphi^I(x,\eta)= \varphi_\infty^I+ \frac{c^I}{2} \ln\!\left(\frac{F(\eta)}{\Omega^2}\right),
\quad
\delta_{IJ}c^Ic^J=2(1-\beta^2).
\end{equation}
In the single canonical-scalar limit this branch overlaps with the accelerating FJNW solution discussed in Ref.~\cite{AnjomshoaMirzaAzizallahi2025}, where an equivalent scalar geometry was also related to a Buchdahl transformation. The additional result here is the derivation as one application of the general cyclic Ricci-source theorem and the systematic extension to arbitrary nonlinear-sigma target geodesics and to their Jordan-frame descendants.
while for a constant target metric $K_{IJ}$,
\begin{equation}\label{eq:cmetric_K}
\varphi^I(x,\eta)=\varphi_\infty^I+\frac{c^I}{2}\ln\!\left(\frac{F(\eta)}{\Omega^2}\right),
\quad
K_{IJ}c^Ic^J=2(1-\beta^2).
\end{equation}

For the hyperbolic target, the vertical branch is
\begin{equation}\label{eq:cmetric_hyperbolic_vertical}
X(x,\eta)=X_0,
\quad
Y(x,\eta)=Y_0\left(\frac{F(\eta)}{\Omega^2}\right)^{\sqrt{2(1-\beta^2)}/(2\ell)} ,
\end{equation}
and the semicircular branch is
\begin{align}\label{eq:cmetric_hyperbolic_semicircle}
X(x,\eta)&=X_c+R\tanh\!\left[u_0+\frac{\sqrt{2(1-\beta^2)}}{2\ell}\ln\!\left(\frac{F(\eta)}{\Omega^2}\right)\right],
\\
Y(x,\eta)&=R\sech\!\left[u_0+\frac{\sqrt{2(1-\beta^2)}}{2\ell}\ln\!\left(\frac{F(\eta)}{\Omega^2}\right)\right].
\end{align}

For the axion--dilaton target, let $a_{\rm ax}$ denote the axion field. The pure dilaton branch is
\begin{equation}\label{eq:cmetric_pure_dilaton}
a_{\rm ax}(x,\eta)=a_0,
\quad
\phi(x,\eta)=\phi_0-\frac{\sqrt{2(1-\beta^2)}}{2}\ln\!\left(\frac{F(\eta)}{\Omega^2}\right),
\end{equation}
For compactness, define
\begin{equation}\label{eq:cmetric_axion_argument}
\Xi(x,\eta)=u_0+\frac{b\sqrt{2(1-\beta^2)}}{2}
\ln\!\left(\frac{F(\eta)}{\Omega^2}\right).
\end{equation}
Then the mixed branch is
\begin{align}\label{eq:cmetric_axion_dilaton}
b\,a_{\rm ax}(x,\eta)&=X_c+R\tanh\Xi(x,\eta),
\\
e^{-b\phi(x,\eta)}&=R\sech\Xi(x,\eta).
\end{align}

For the compact spherical target, the meridian branch is
\begin{align}\label{eq:cmetric_spherical_meridian}
\Theta(x,\eta)
&=\Theta_0+\frac{\sqrt{2(1-\beta^2)}}{2\ell}
\ln\!\left(\frac{F(\eta)}{\Omega^2}\right),
\\
\Phi_a(x,\eta)&=\Phi_{a0}.
\end{align}
and the two-coordinate great-circle branch is
\begin{align}\label{eq:cmetric_spherical_general}
s(x,\eta)&=s_0+\frac{\sqrt{2(1-\beta^2)}}{2\ell}\ln\!\left(\frac{F(\eta)}{\Omega^2}\right),
\\
\cos\Theta(x,\eta)
&=\sin\alpha\,\sin s(x,\eta),
\\
\tan\!\left[\Phi_a(x,\eta)-\Phi_{a0}\right]&=\cos\alpha\,\tan s(x,\eta).
\end{align}

The displayed family is an accelerating Buchdahl--sigma solution on the static patch specified above. Other C-metric regions in which the causal character of the chosen cyclic direction changes must be treated patchwise, with the corresponding fixed sign $\eps$ and $\Sigma=\frac12\ln|g_{yy}^{(0)}|$. The Buchdahl potential on the present static patch is
\begin{equation}\label{eq:cmetric_buchdahl_potential}
\Sigma(x,\eta)=\frac12\ln\!\left(\frac{F(\eta)}{\Omega^2}\right),
\end{equation}
so the scalar map is controlled by the C-metric redshift factor and by one target-space geodesic.

The axis structure is still governed locally by the roots of $G(x)$. Since the $(x,\phi)$ block is multiplied by a common Buchdahl conformal factor, the local conical regularity condition at a simple axis root $x_i$ remains
\begin{equation}\label{eq:phys_cmetric_axis_period}
\Delta\phi_i = \frac{4\pi}{|G'(x_i)|},
\end{equation}
up to the usual choice of which axis segment is regularized. The scalar dressing therefore does not remove the string/strut imbalance of the vacuum C-metric. It is hence natural to ask if the conclusion could be changed by choosing the azimuthal coordinate itself as the Buchdahl direction. The answer is no within the present construction. Near a simple
axis root $x=x_i$ of $G(x)$,
\begin{equation}
G(x)
\simeq
G'(x_i)(x-x_i).
\end{equation}
If one chooses $y=\phi$ in $D=4$, the generated metric has, up to finite nonzero factors,
\begin{equation}\label{eq:cmetric_phi_axis_scaling}
g_{\phi\phi}^{(\beta)}
\sim
G^\beta,
\qquad
g_{xx}^{(\beta)}
\sim
G^{-\beta}.
\end{equation}
The proper radial distance from the axis consequently behaves as
\begin{equation}\label{eq:cmetric_phi_axis_rho}
\rho
\sim
\int |x-x_i|^{-\beta/2}\,\dd x
\sim
|x-x_i|^{1-\beta/2},
\end{equation}
whereas the azimuthal circumference behaves as
\begin{equation}\label{eq:cmetric_phi_axis_circumference}
C_\phi\sim|x-x_i|^{\beta/2}.
\end{equation}
Elementary flatness requires
\begin{equation}
C_\phi\propto\rho ,
\end{equation}
and comparison of the two powers gives
\begin{equation}
\frac{\beta}{2}=1-\frac{\beta}{2},
\quad\Longrightarrow\qquad
\beta=1 .
\end{equation}
For a positive-definite target this is precisely the zero-scalar-speed seed endpoint. Hence a nontrivial Buchdahl scalar deformation cannot remove the C-metric conical imbalance. With the temporal cyclic direction used above, the two deficit angles are unchanged, whereas with the azimuthal cyclic direction the axis fails to be elementary-flat unless the deformation is trivial.

A regular accelerating scalar geometry free of the string/strut imbalance would therefore require additional structure beyond the present rank-one Buchdahl deformation, for example additional matter or external fields, or a more general solution-generating
construction. What changes is the nature of the surfaces $F(\eta)=0$. There $\Sigma$ diverges logarithmically. In the generated metric one has
\begin{equation}\label{eq:phys_cmetric_inverse_scaling}
g_\beta^{\eta\eta}=\Omega^{4-2\beta}F^\beta,
\qquad
g_\beta^{xx}=\Omega^{4-2\beta}G\,F^{\beta-1}.
\end{equation}
Moreover,
\begin{equation}
\partial_\eta\Sigma=\frac{F'}{2F}-\partial_\eta\ln\Omega,
\qquad \partial_x\Sigma=-\partial_x\ln\Omega .
\end{equation}

Therefore, near a simple root $\eta=\eta_0$ of $F$, with $F'(\eta_0)\neq0$, and with $G$, $\Omega$ finite and nonzero, the $\eta$-derivative gives the leading contribution, \begin{equation}\label{eq:phys_cmetric_horizon_behavior}
R[g_\beta] = C_4(\beta)(\nabla\Sigma)^2\sim C_4(\beta)\,\frac{\Omega^{4-2\beta}F'(\eta_0)^2}{4}\,
F^{\beta-2}.
\end{equation}

The $x$-derivative contribution is subleading, behaving as $F^{\beta-1}$. Hence, for the nontrivial positive-target branch $-1<\beta<1$, the would-be acceleration or black-hole horizon of the seed is generically converted into a scalar curvature singularity.

\subsubsection{Jordan-frame descendants of the new families}

All Einstein-frame solutions above immediately generate tensor--multiscalar solutions in the Jordan frame, using the standard Einstein/Jordan-frame dictionary of tensor--multiscalar gravity~\cite{Damour:1992we}. If
\begin{equation}\label{eq:new_examples_jordan_conformal}
g_{AB} = \mathcal A(\varphi)^{2/(D-2)}\tilde g_{AB},
\end{equation}
then the Jordan-frame line element is
\begin{equation}\label{eq:new_examples_jordan_general}
\dd\tilde s_\beta^2=\mathcal A(\Phi(\Sigma))^{-2/(D-2)}\dd s_\beta^2.
\end{equation}

For example, for the scalar Kaluza--Klein bubble \eqref{eq:kk_bubble_generated} and a two-field coupling
\begin{equation}\label{eq:new_examples_A_example}
\mathcal A(\phi,a)=\exp(2\alpha\phi+2\delta a),
\end{equation}
one obtains
\begin{align}\label{eq:kk_bubble_jordan_example}
\dd\tilde s_\beta^2&= \exp\!\left[ -\frac{4}{D-2}\left(\alpha\phi(r)+\delta a(r)\right) \right] \Bigg\{ F^\beta \dd\psi^2 \nonumber\\
&+F^{(1-\beta)/(D-3)} \left[ -\dd t^2 + F^{-1}\dd r^2 + r^2\dd\Omega_{D-3}^2 \right] \Bigg\}.
\end{align}
Two target geodesics with the same speed give the same Einstein-frame metric, but they generally give different Jordan-frame metrics because $\mathcal A(\Phi(\Sigma))$ depends on the position of the geodesic in target space. This is especially important for the bubble, Rosen-wave and Gowdy families: their Einstein-frame metrics are generated by the same rank-one source mechanism, while their Jordan-frame geometries can differ in bubble radius, wave amplitude, cosmological scale factors and effective matter-frame singularity structure.

The main lesson of these examples is that the Buchdahl--sigma generator is not only a scalar-dressing of black-hole seeds. It also generates spacelike-cyclic bubble geometries, null-gradient wave solutions, inhomogeneous compact cosmologies and accelerating target-geodesic families. The bubble, Rosen and Gowdy cases are particularly clean applications because they use cyclic directions and potentials that are not the standard static radial redshift potential. They therefore make the target-covariant content of the theorem visible in settings where the scalar source is respectively spacelike, null-gradient and cosmological.

\section{Conclusion}
\label{sec:interpretation}

We have developed a Buchdahl solution generator for Einstein gravity coupled to nonlinear sigma models. The central result is a geometric Ricci-source theorem: on any connected region where a Ricci-flat seed admits a non-null hypersurface-orthogonal cyclic coordinate with fixed-sign nonzero norm, the cyclic geometry determines a universal rank-one source whose strength depends only on the spacetime dimension and the Buchdahl parameter. Composing the associated Buchdahl potential with an affinely parametrized geodesic of the target metric then yields an exact solution of the full Einstein--sigma equations. For positive-definite targets, the rank-one pullback condition implies that the nontrivial scalar map has one-dimensional image. On regions where the Buchdahl gradient is non-null, the scalar equations further require this curve to be geodesic within the ansatz $\varphi^I=\Phi^I(\Sigma)$, whereas the null-gradient sector admits the broader constant-speed class discussed above. For indefinite targets, additional degenerate realizations associated with null target directions may also exist.
	
The main conceptual advantage of the construction is the clean separation between spacetime geometry and target-space dynamics. The Ricci-flat cyclic seed determines the Buchdahl potential and the resulting invariant source, whereas the internal scalar structure is fixed independently by a geodesic of the chosen target manifold. The generator can therefore be viewed as a three-step procedure: a Ricci-flat cyclic seed determines the Buchdahl potential, the Buchdahl deformation produces a universal rank-one Ricci source, and this source is realized by an affinely parametrized target-space geodesic. This provides a unified framework encompassing canonical scalars, curved nonlinear sigma models and vacuum tensor--multiscalar descendants.
	
The explicit examples show that the mechanism extends far beyond the static spherical sector. Tangherlini and black-brane seeds generate higher-dimensional scalar-dressed geometries, while Weyl solutions transfer the multipolar structure of the vacuum seed directly to the sigma-model sector. Kasner seeds yield scalar--Kasner cosmologies, and spacelike-cyclic Kaluza--Klein bubbles, null-gradient Rosen waves, polarized Gowdy cosmologies and accelerating C-metric configurations demonstrate that the Buchdahl potential may be radial, cosmological, inhomogeneous or null-gradient. In the familiar static black-hole and bubble examples, a nonzero positive-target scalar speed generically converts a regular zero of the seed cyclic norm into a curvature singularity, whereas the wave and cosmological examples exhibit qualitatively different realizations of the same underlying geometric mechanism.

The horizon analysis done also emphasizes the physical
interpretation of the static families. For positive-definite targets, the nontrivial Tangherlini-derived branch is not a family of scalar-hairy black holes: the Schwarzschild--Tangherlini horizon is converted into a curvature singularity. Likewise, a nontrivial Buchdahl deformation of the Kaluza--Klein bubble destroys the smooth bubble surface. The accelerating C-metric examples exhibit the same generic mechanism at simple zeros of the seed redshift function. Consequently, the scalar charges carried by these nontrivial branches should be interpreted as asymptotic charges of exact scalar geometries, rather than as thermodynamic scalar hair attached to a regular black-hole horizon. Within the positive-definite Tangherlini family, the only regular Schwarzschild--Tangherlini black-hole endpoint is $\beta=1$, where both the scalar speed and the scalar charge vanish.
	
The Jordan-frame extension further enriches the construction. Target geodesics with the same invariant speed generate the same Einstein-frame geometry but can lead to inequivalent Jordan-frame metrics because the conformal factor depends on the position of the trajectory in target space. A single cyclic vacuum seed may therefore generate families that are geometrically equivalent in the Einstein frame while remaining physically distinct in the matter frame.
	
Several extensions are natural. A particularly interesting direction is the construction of higher-rank generators based on several independent cyclic potentials, in which the present rank-one source would be replaced by a matrix-valued combination of spacetime gradients and the target-space geodesic by an appropriate higher-dimensional totally geodesic submanifold. A complementary direction is the systematic study of the null-gradient sector identified here, exemplified by the Rosen-wave family, where the scalar equations admit more general constant-speed target curves beyond the geodesic subclass considered explicitly in this work. Further developments include non-orthogonal Killing sectors, gauge fields and scalar potentials, together with systematic studies of global structure, conserved charges, stability and singularity formation in the non-spherical, cosmological, wave and accelerating families. The rank-one construction established here provides a simple and broadly applicable foundation for these generalizations.

\begin{acknowledgments}
We thank the anonymous referee for the valuable suggestions provided, which have greatly contributed to enhancing the quality of this manuscript.
The authors acknowledge funding from the Fundação para a Ciência e a Tecnologia (FCT) through the research grant UID/04434/2025.
FSNL also acknowledges support from the FCT Scientific Employment Stimulus contract with reference CEECINST/00032/2018.
\end{acknowledgments}

\appendix

\section{Cyclic curvature-density identity}
\label{app:cyclic_density_derivation}

This appendix derives the geometric identity used in Sec.~\ref{sec:method}. The calculation is local and does not assume any field equation. We start from
\begin{equation}\label{app:eq:cyclic_metric}
\dd s_D^2 = \eps e^{2\gamma(x)}\dd y^2 + e^{2\chi(x)}h_{ij}(x)\dd x^i\dd x^j ,
\end{equation}
where all fields are independent of $y$. It is useful to factor out the conformal scale $e^{2\chi}$ and write
\begin{equation}
g_{AB}=e^{2\chi}\widetilde g_{AB},
\quad
\dd \widetilde s^2= \eps e^{2u}\dd y^2+ h_{ij}\dd x^i\dd x^j ,
\quad
u=\gamma-\chi .
\end{equation}
For a one-dimensional warped fibre, the Ricci scalar of
$\widetilde g_{AB}$ is
\begin{equation}\label{app:eq:warped_ricci}
R[\widetilde g]= R[h]-2\D^2u-2(\D u)^2 ,
\end{equation}
where $\D_i$ is the covariant derivative of $h_{ij}$. The conformal transformation in $D$ dimensions gives
\begin{align}\label{app:eq:conformal_ricci}
R[g]=e^{-2\chi}\Big[&R[\widetilde g]-2(D-1)\widetilde\Box\chi
\nonumber\\
&-(D-1)(D-2)(\widetilde\nabla\chi)^2\Big].
\end{align}

Since $\chi$ is independent of $y$,
\begin{equation}
(\widetilde\nabla\chi)^2=(\D\chi)^2,
\quad
\widetilde\Box\chi=\D^2\chi+\D u\cdot\D\chi .
\end{equation}
Using $u=\gamma-\chi$, Eqs.~\eqref{app:eq:warped_ricci} and
\eqref{app:eq:conformal_ricci} yield
\begin{align}\label{app:eq:raw_ricci_cyclic}
R[g]&=e^{-2\chi}\Big\{R[h]-2\D^2\gamma-2(D-2)\D^2\chi-2(\D\gamma)^2\nonumber\\
&-2(D-3)\D\gamma\cdot\D\chi-(D-2)(D-3)(\D\chi)^2\Big\}.
\end{align}
The determinant is
\begin{equation}
\sqrt{|g_D|}=e^{\gamma+(D-1)\chi}\sqrt{|h|}.
\end{equation}
Therefore
\begin{align}\label{app:eq:raw_density_cyclic}
&\sqrt{|g_D|}R[g]=\sqrt{|h|}e^F\Big\{ R[h] - 2\D^2\gamma - 2(D-2)\D^2\chi\nonumber\\
&- 2(\D\gamma)^2-2(D-3)\D\gamma\cdot\D\chi - (D-2)(D-3)(\D\chi)^2 \Big\},
\end{align}
where
\begin{equation}
F=\gamma+(D-3)\chi .
\end{equation}

We now remove total divergences on the base. For any scalar $X$,
\begin{equation}
\sqrt{|h|}e^F\D^2X=\partial_i\!\left(\sqrt{|h|}e^F\D^iX\right)-\sqrt{|h|}e^F\D F\cdot\D X .
\end{equation}
Dropping the divergence terms gives
\begin{align}
-2e^F\D^2\gamma
&\doteq
2e^F\D F\cdot\D\gamma ,
\\
-2(D-2)e^F\D^2\chi
&\doteq
2(D-2)e^F\D F\cdot\D\chi .
\end{align}
Since
\begin{equation}
\D F=\D\gamma+(D-3)\D\chi ,
\end{equation}
the gradient terms combine as
\begin{align}
&2\D F\cdot\D\gamma+2(D-2)\D F\cdot\D\chi-2(\D\gamma)^2\nonumber\\
&\quad-2(D-3)\D\gamma\cdot\D\chi-(D-2)(D-3)(\D\chi)^2\nonumber\\
&=(D-2)\left[2\D\gamma\cdot\D\chi+(D-3)(\D\chi)^2\right].
\end{align}
Hence
\begin{align}\label{app:eq:reduced_density_cyclic}
\sqrt{|g_D|}R[g_D]
&\doteq
\sqrt{|h|}e^F
\left\{R[h]+(D-2)\left[
2\D\gamma\cdot\D\chi
\right.\right.\nonumber\\
&\left.\left.+(D-3)(\D\chi)^2\right]\right\}.
\end{align}
This is Eq.~\eqref{eq:D_density_general}. Notice in particular the coefficient $-2(D-3)$ of the mixed term in Eq.~\eqref{app:eq:raw_ricci_cyclic}; this is the coefficient required for the reduced density \eqref{app:eq:reduced_density_cyclic}.

We now specialize the off-shell identity \eqref{app:eq:reduced_density_cyclic} to the Buchdahl curve
\begin{equation}
\gamma=\beta\Sigma,
\quad
\chi=\frac{1-\beta}{D-3}\Sigma .
\end{equation}
Then
\begin{equation}
F=\gamma+(D-3)\chi=\beta\Sigma+(1-\beta)\Sigma=\Sigma .
\end{equation}
Moreover,
\begin{align}
&2\D\gamma\cdot\D\chi+(D-3)(\D\chi)^2=2\beta\frac{1-\beta}{D-3}(\D\Sigma)^2
\nonumber\\
&\quad+ (D-3)\frac{(1-\beta)^2}{(D-3)^2}(\D\Sigma)^2
=\frac{1-\beta^2}{D-3}(\D\Sigma)^2 .
\end{align}
Therefore
\begin{equation}
\sqrt{|g_\beta|}R[g_\beta]
\doteq
\sqrt{|h|}e^\Sigma
\left[
R[h]
+
\frac{D-2}{D-3}(1-\beta^2)(\D\Sigma)^2
\right].
\end{equation}
For $\beta=1$, one has $\chi=0$, $\gamma=\Sigma$, and hence
\begin{equation}
\sqrt{|g_1|}R[g_1]
\doteq
\sqrt{|h|}e^\Sigma R[h].
\end{equation}
Subtracting the two densities gives
\begin{equation}
\sqrt{|g_\beta|}R[g_\beta]
-
\sqrt{|g_1|}R[g_1]
\doteq
\sqrt{|h|}e^\Sigma
\frac{D-2}{D-3}(1-\beta^2)(\D\Sigma)^2 .
\end{equation}
This is the geometric origin of the effective scalar kinetic term.

\section{Direct Ricci form of the Buchdahl system and field equations check}\label{app:field_equation_check}

The reduced-density identity is useful for deriving the coefficient $C_D(\beta)$. The main text already gives the concise Ricci-level proof of the solution-generating theorem; here we provide the detailed component calculation and the unrestricted field-equation check. Assume that the $\beta=1$ member
\begin{equation}
\dd s_0^2
=
\eps e^{2\Sigma}\dd y^2+h_{ij}\dd x^i\dd x^j
\end{equation}
is Ricci-flat. Its nonzero Ricci equations are equivalent to
\begin{align}
\D^2\Sigma+(\D\Sigma)^2&=0,
\label{eq:seed_sigma_equation}\\
R_{ij}[h]&=\D_i\D_j\Sigma+\D_i\Sigma\D_j\Sigma,
\label{eq:seed_base_equation}
\end{align}
where
\begin{equation}
\D^2\Sigma=h^{ij}\D_i\D_j\Sigma,
\quad
(\D\Sigma)^2=h^{ij}\D_i\Sigma\D_j\Sigma .
\end{equation}

Now set
\begin{equation}
a=\frac{1-\beta}{D-3},
\quad
k=\beta-a .
\end{equation}
The Buchdahl metric may be written as
\begin{equation}
\dd s_\beta^2
=
e^{2a\Sigma}
\left[
\eps e^{2k\Sigma}\dd y^2+h_{ij}\dd x^i\dd x^j
\right].
\end{equation}
A direct Ricci calculation gives
\begin{equation}
R_{yy}[g_\beta]
=
-\eps e^{2k\Sigma}\,
\beta\left[
\D^2\Sigma+(\D\Sigma)^2
\right],
\end{equation}
and hence, by Eq.~\eqref{eq:seed_sigma_equation},
\begin{equation}
R_{yy}[g_\beta]=0 .
\end{equation}
For the base components one obtains
\begin{align}
R_{ij}[g_\beta]
&=R_{ij}[h]-\D_i\D_j\Sigma
\nonumber\\
&\quad+\left[-k^2+(D-2)a^2\right]
\D_i\Sigma\D_j\Sigma
\nonumber\\
&\quad-a h_{ij}\left[\D^2\Sigma+(\D\Sigma)^2\right].
\end{align}

Using Eqs.~\eqref{eq:seed_sigma_equation} and
\eqref{eq:seed_base_equation}, this reduces to
\begin{equation}
R_{ij}[g_\beta]
=
\left[
1-k^2+(D-2)a^2
\right]
\D_i\Sigma\D_j\Sigma .
\end{equation}
Since
\begin{equation}
1-k^2+(D-2)a^2
=
\frac{D-2}{D-3}(1-\beta^2),
\end{equation}
we find
\begin{equation}
R_{ij}[g_\beta]
=
C_D(\beta)\D_i\Sigma\D_j\Sigma .
\end{equation}
Combining the $yy$ and base components,
\begin{equation}\label{eq:Ricci_Buchdahl_direct}
R_{AB}[g_\beta]
=
C_D(\beta)\partial_A\Sigma\partial_B\Sigma .
\end{equation}

The same calculation also gives the wave equation for the Buchdahl potential. Indeed,
\begin{align}
\Box_{g_\beta}\Sigma
&=
e^{-2a\Sigma}
\left[
\D^2\Sigma+\big(k+(D-2)a\big)(\D\Sigma)^2
\right]
\nonumber\\
&=e^{-2a\Sigma}\left[\D^2\Sigma+(\D\Sigma)^2\right],
\end{align}
because $k+(D-2)a=1$. Therefore
\begin{equation}\label{eq:Sigma_wave_direct}
\Box_{g_\beta}\Sigma=0 .
\end{equation}

Equations~\eqref{eq:Ricci_Buchdahl_direct} and \eqref{eq:Sigma_wave_direct} are the effective one-scalar Buchdahl system generated from a Ricci-flat cyclic seed.

\subsection{Field-equation check}

We now verify directly that the unrestricted Einstein--sigma field equations are satisfied. The physical field equations are
\begin{equation}
R_{AB}[g]
=
\G_{IJ}(\varphi)\partial_A\varphi^I\partial_B\varphi^J,
\end{equation}
and
\begin{equation}
\Box_g\varphi^I
+
\Gamma^I{}_{JK}(\varphi)
g^{AB}\partial_A\varphi^J\partial_B\varphi^K
=
0.
\end{equation}
From the Buchdahl construction, the metric $g_\beta$ satisfies
\begin{equation}
R_{AB}[g_\beta]
=
C_D(\beta)\partial_A\Sigma\partial_B\Sigma,
\quad
\Box_{g_\beta}\Sigma=0.
\end{equation}
Let
\begin{equation}
\varphi^I(x)=\Phi^I(\Sigma(x)).
\end{equation}
Then
\begin{equation}
\partial_A\varphi^I
=
\frac{\dd\Phi^I}{\dd\Sigma}\partial_A\Sigma .
\end{equation}
Therefore
\begin{equation}
\G_{IJ}(\varphi)\partial_A\varphi^I\partial_B\varphi^J
=
\left[
\G_{IJ}(\Phi)
\frac{\dd\Phi^I}{\dd\Sigma}
\frac{\dd\Phi^J}{\dd\Sigma}
\right]
\partial_A\Sigma\partial_B\Sigma .
\end{equation}
Using the speed condition
\begin{equation}
\G_{IJ}(\Phi)
\frac{\dd\Phi^I}{\dd\Sigma}
\frac{\dd\Phi^J}{\dd\Sigma}
=
C_D(\beta),
\end{equation}
we obtain
\begin{equation}
\G_{IJ}(\varphi)\partial_A\varphi^I\partial_B\varphi^J
=
C_D(\beta)\partial_A\Sigma\partial_B\Sigma
=
R_{AB}[g_\beta].
\end{equation}
Hence the metric equations are satisfied. For the scalar equations, first note that
\begin{equation}
\Box_{g_\beta}\varphi^I=\frac{\dd\Phi^I}{\dd\Sigma}\Box_{g_\beta}\Sigma+\frac{\dd^2\Phi^I}{\dd\Sigma^2}g_\beta^{AB}\partial_A\Sigma\partial_B\Sigma .
\end{equation}
Moreover,
\begin{equation}
g_\beta^{AB}\partial_A\varphi^J\partial_B\varphi^K
=
\frac{\dd\Phi^J}{\dd\Sigma}
\frac{\dd\Phi^K}{\dd\Sigma}
g_\beta^{AB}\partial_A\Sigma\partial_B\Sigma .
\end{equation}
Therefore
\begin{align}
&\Box_{g_\beta}\varphi^I
+\Gamma^I{}_{JK}(\Phi)
 g_\beta^{AB}\partial_A\varphi^J\partial_B\varphi^K
=\frac{\dd\Phi^I}{\dd\Sigma}\Box_{g_\beta}\Sigma
\nonumber\\
&\quad+\left[
\frac{\dd^2\Phi^I}{\dd\Sigma^2}
+\Gamma^I{}_{JK}(\Phi)
\frac{\dd\Phi^J}{\dd\Sigma}
\frac{\dd\Phi^K}{\dd\Sigma}
\right]
(\nabla\Sigma)^2 .
\end{align}

Using
\begin{equation}
\Box_{g_\beta}\Sigma=0
\end{equation}
and the target-space geodesic equation
\begin{equation}
\frac{\dd^2\Phi^I}{\dd\Sigma^2}+\Gamma^I{}_{JK}(\Phi)\frac{\dd\Phi^J}{\dd\Sigma}\frac{\dd\Phi^K}{\dd\Sigma}=0,
\end{equation}
the scalar equations vanish identically. Thus the full Einstein--sigma equations are satisfied.

\bibliographystyle{apsrev4-2}
\bibliography{apssamp}

\end{document}